\documentclass{article} %
\usepackage{iclr2027_conference,times}

\usepackage[utf8]{inputenc}
\usepackage[T1]{fontenc}
\usepackage{hyperref}
\usepackage{url}
\usepackage{booktabs}
\usepackage{amsfonts}
\usepackage{amsmath}
\usepackage{nicefrac}
\usepackage{microtype}
\usepackage{xcolor}   %
\usepackage{graphicx}
\usepackage{placeins}
\usepackage{multirow}
\usepackage{algorithm}
\usepackage{algorithmic}
\usepackage{float}
\usepackage{wrapfig}
\usepackage{etoolbox}
\newcommand{\indicator}{\mathbf{1}}
\usepackage{capt-of}

\AtBeginEnvironment{table}{%
  \setlength{\abovecaptionskip}{0pt}%
  \setlength{\belowcaptionskip}{7pt}}

\newcommand{\tablefootnotesize}{\footnotesize}

\title{\fontsize{16pt}{19pt}\selectfont WeaveMark: Robust and Scalable Multi-bit \\ LLM Watermarking via Coded Payload Spreading}

\author{Gang-Hyun Park \quad Ju-Hyeong Lee \quad Hee-Youl Kwak \quad Dae-Young Yun \\
University of Ulsan, Ulsan, Republic of Korea \\
\texttt{\{qkrrkd90, xkca2446, ghy1228, dyyun95\}@gmail.com}}

\iclrfinalcopy

\begin{document}

\raggedbottom

\maketitle

\begin{abstract}
Multi-bit watermarking for large language models enables content source tracing by embedding user-identifiable messages into generated text. Existing methods face a fundamental trade-off among extraction accuracy, text quality, and payload capacity. We propose \textsc{WeaveMark}\footnote{Code is available at \url{https://anonymous.4open.science/r/WeaveMark-ED6F}.}, a robust and scalable multi-bit LLM watermarking scheme based on coded payload spreading. \textsc{WeaveMark} shifts this trade-off frontier by improving payload capacity through multi-bit-per-token spreading (weaving), improving extraction accuracy through soft-decision error-correcting codes, and preserving text quality through unbiased multilayer reweighting. It further introduces dedicated zero-bit layers for reliable watermark presence detection. Extensive experiments demonstrate substantial gains in extraction performance, especially for long messages and edited text, without degrading text quality. \textsc{WeaveMark} achieves an 89.8\% match rate for 32-bit messages at 200 tokens, compared with 20.8\% for BiMark. Under 10\% substitution attacks on 16-bit messages at 200 tokens, it maintains 86.0\% versus 30.7\%.
\end{abstract}

\section{Introduction}

Large language models (LLMs) such as GPT~\citep{openai2023gpt4} and LLaMA~\citep{dubey2024llama3} are widely used in chatbots, search engines, and programming assistants. However, they can also be misused to generate fake news~\citep{zellers2019defending}, phishing emails, and fraudulent reviews, raising concerns about LLM safety~\citep{weidinger2022taxonomy}. Watermarking addresses this risk by embedding traceable information into generated text for provenance verification and source tracing~\citep{kirchenbauer2023watermark, kuditipudi2024robust, christ2024undetectable, hu2024unbiased, dathathri2024synthid}.

Multi-bit watermarking embeds a $k$-bit message, such as a user ID, into generated text for source tracing. Each bit is assigned to generated tokens and extracted from token-level statistics. This creates a fundamental trade-off among \emph{extraction accuracy}, \emph{text quality}, and \emph{payload capacity}. Improving extraction accuracy requires more distinguishable token-level statistics, which can perturb the original token distribution and degrade text quality. Meanwhile, increasing payload capacity leaves fewer token-level observations per bit, making reliable extraction harder. A practical method must therefore recover messages accurately, preserve the original text distribution, and support large payloads.

Existing methods occupy different points on this trade-off frontier. BiMark~\citep{feng2025bimark} preserves text quality through multilayer unbiased reweighting, but its one-bit-per-token design limits payload capacity. 
In contrast, MPAC~\citep{yoo2024mpac} and RSBH~\citep{qu2025provably} improve payload capacity through segment-based embedding, but rely on biased reweighting that degrades text quality.

\begin{figure}[t]
\centering
\includegraphics[width=\textwidth]{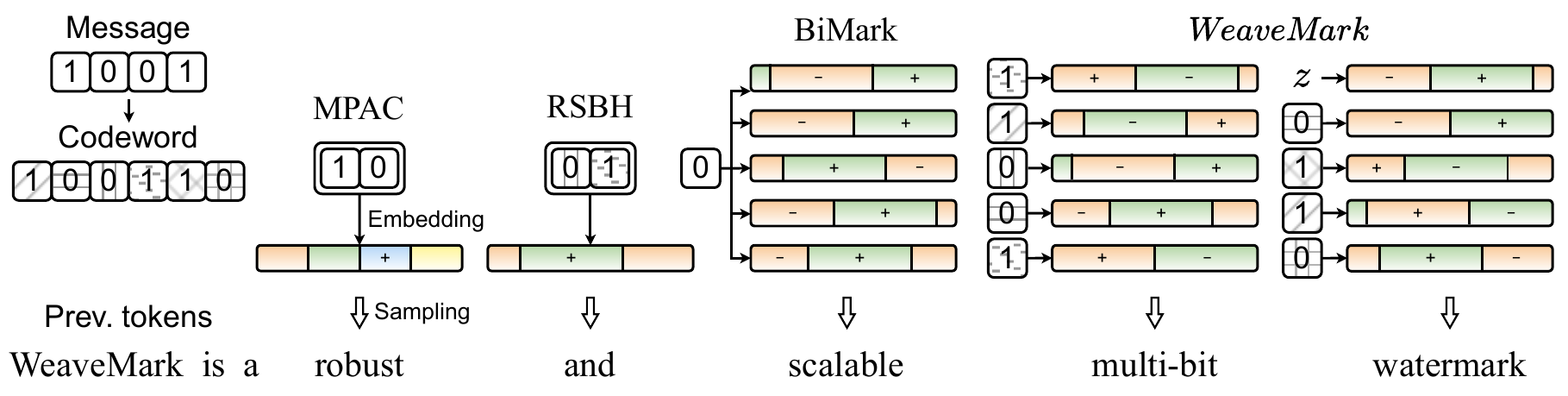}
\vspace{-7pt}
\caption{Conceptual comparison of representative multi-bit LLM watermarking methods. Each column represents a distinct scheme. Colors denote vocabulary partitions, $+$/$-$ signs indicate reweighting directions, and hatched bits denote ECC codeword bits. \textsc{WeaveMark} weaves multiple codeword bits into each token across layers and uses soft-decision ECC decoding.}
\label{fig:scheme_diagram}
\end{figure}

In this work, we propose \textsc{WeaveMark}, a robust and scalable multi-bit LLM watermarking scheme based on coded payload spreading.
The name reflects the idea of weaving ECC-coded payload bits across generated tokens and watermarking layers.
\textsc{WeaveMark} increases payload capacity through coded payload spreading, improves extraction accuracy through soft-decision error-correcting code (ECC), and preserves text quality through unbiased multilayer reweighting.
Together, these components shift the fundamental trade-off frontier of multi-bit watermarking.
Figure~\ref{fig:scheme_diagram} compares \textsc{WeaveMark} with representative prior methods~\citep{yoo2024mpac,qu2025provably,feng2025bimark}.

Practical systems also need zero-bit detection. Because the target partition depends on the unknown message, existing multi-bit schemes use the most-voted partition as a proxy, which inflates the null statistic and raises the detection threshold. \textsc{WeaveMark} avoids this with message-independent zero-bit layers.

Our contributions are summarized as follows:
\begin{enumerate}
\item \textbf{Trade-off analysis.} We clarify the trade-off relationship among existing multi-bit LLM watermarking schemes in terms of extraction accuracy, text quality, and payload capacity.

\item \textbf{Improved payload capacity via coded payload spreading.} We propose coded payload spreading, which combines multi-bit-per-token embedding and layer shuffling to distribute token-level statistics and layer reliability more evenly across coded bits. This enables longer payloads while retaining unbiased reweighting.

\item \textbf{Strong multi-bit message recovery.} We combine payload spreading with soft-decision ECC decoding and show that \textsc{WeaveMark} substantially improves extraction accuracy over state-of-the-art multi-bit watermarking methods, especially for long messages, short generated texts, and editing attacks, while preserving text quality.

\item \textbf{Reliable zero-bit detection.} We introduce dedicated zero-bit layers that avoid the max-over-partitions effect, enabling reliable zero-bit detection while preserving multi-bit recovery.
\end{enumerate}

\section{Related work}
\label{sec:llm_watermarking}

\textbf{Zero-bit watermarking.}
Zero-bit watermarking detects only whether a text was generated by a watermarked model. The green-list watermark of \citet{kirchenbauer2023watermark} uses the preceding tokens to pseudorandomly split the vocabulary into a green list and a red list, biases sampling toward the green list, and detects the watermark by testing for an unusually large number of green-list tokens.

\textbf{Multi-bit watermarking.}
Multi-bit watermarking assigns message bits to generated tokens and aggregates token-level statistics to recover each bit. Because the message is recovered only when every bit is correct, extraction becomes harder as the message length grows.

Existing methods take different approaches to this challenge in how they map message bits to vocabulary partitions. MPAC~\citep{yoo2024mpac} assigns each token to a message segment and embeds the corresponding segment value by biasing one of multiple vocabulary partitions. This segment-allocation strategy improves payload capacity, but increasing capacity requires larger segments and therefore smaller vocabulary partitions, which degrades text quality by constraining token selection.

RSBH~\citep{qu2025provably} takes a different approach by combining binary vocabulary partitioning with segment allocation. Binary partitioning provides more candidate tokens per partition, reducing quality degradation compared with MPAC. However, recovering multi-valued segment information from binary partitions increases bit-level errors. It uses Reed-Solomon ECC to mitigate these errors.

To preserve text quality, BiMark~\citep{feng2025bimark} takes another direction through multilayer unbiased reweighting. Each layer uses an independent vocabulary bipartition, allowing BiMark to preserve the expected token distribution. Compared with segment-wise embedding, its bit-wise embedding enables more reliable bit-level extraction. However, since each token is assigned to only one message bit, long messages and short generated texts receive too few and unevenly distributed votes per bit, limiting payload capacity.

\begin{wraptable}{r}{0.48\textwidth}
\vspace{-1.5em}
\centering
\caption{Multi-bit watermarking comparison}
\label{tab:scheme_properties}
\vspace{0.3em}
\small
\setlength{\tabcolsep}{4pt}
\renewcommand{\arraystretch}{1.08}
\begin{tabular}{lccc}
\toprule
Method & \shortstack{Extraction\\accuracy$^{*}$} & \shortstack{Text\\quality$^{\dagger}$} & \shortstack{Payload\\capacity$^{\diamond}$} \\
\midrule 
MPAC {\tablefootnotesize($\delta{=}3.0$)} & 60.5 & 26.5 & < 12 \\
RSBH {\tablefootnotesize($\delta{=}3.0$)} & 83.5 & 27.5 & 24 \\
BiMark & 83.0 & 30.3 & 16 \\
\textsc{WeaveMark} & \textbf{96.6} & \textbf{30.5} & \textbf{> 32} \\
\bottomrule
\end{tabular}
\vspace{0.3em}
\raggedright
\scriptsize 
\\
$^{*}$ Match rate (\%) for 16-bit messages with 200 generated tokens.\\
$^{\dagger}$ BERTScore on summarization (CNN/DailyMail).\\
$^{\diamond}$ Max message length achieving 80\% match rate with 200 tokens.
\vspace{-1.2em}
\end{wraptable}

\textbf{Trade-off summary.}
Existing methods occupy different points on the multi-bit watermarking trade-off frontier. MPAC and RSBH improve payload capacity and extraction accuracy through biased reweighting at the cost of distribution preservation, whereas BiMark preserves text quality through unbiased reweighting but embeds only one bit per token, limiting payload capacity. Table~\ref{tab:scheme_properties} shows that \textsc{WeaveMark} achieves the strongest trade-off across extraction accuracy, text quality, and payload capacity. Appendix~\ref{app:extended} additionally compares against XMark~\citep{xu2026xmark} and StealthInk~\citep{jiang2025stealthink}.

\section{Preliminary}
\label{sec:preliminary}

\textbf{LLM watermarking and zero-bit detection.}
Let $\mathcal{V}$ be the vocabulary and $\Delta_{\mathcal{V}}$ be the set of probability distributions over $\mathcal{V}$. At token position $t$, an autoregressive LLM $M$ defines the original next-token distribution $P_M(\cdot \mid x_{<t}) \in \Delta_{\mathcal{V}}$. \citet{kirchenbauer2023watermark} use the preceding $h$ tokens, denoted by $x_{-h}=x_{t-h:t-1}$, together with a secret key to partition the vocabulary into green-list and red-list sets, and add a bias toward green-list tokens. This can be written as a reweighting function $R:\Delta_{\mathcal{V}}\rightarrow\Delta_{\mathcal{V}}$, which transforms the original distribution into the watermarked distribution $P_{M,w}(\cdot \mid x_{<t}) = R\!\left(P_M(\cdot \mid x_{<t})\right)$, from which the next token is sampled.

Given a text with $T$ tokens, let $|s|_G$ be the number of green-list tokens and $\gamma$ be the green-list proportion. The standard zero-bit detection statistic is $z = (|s|_G-\gamma T)/\sqrt{\gamma(1-\gamma)T}$, and a text is classified as watermarked if $z$ exceeds a threshold.

Multi-bit watermarking embeds a $k$-bit message 
$\mathbf{m}=(m_1,\ldots,m_k)\in\{0,1\}^k$ into generated text. When ECC is used, the message is first encoded into an $n$-bit codeword 
$\mathbf{c}=E(\mathbf{m})\in\{0,1\}^n$, and the codeword bits are embedded instead. The additional $n-k$ bits enable error correction, but also increase the number of bits that must be embedded.
When ECC is not used, we set $\mathbf{c}=\mathbf{m}$ and $n=k$.

\textbf{Unbiased multilayer watermarking.}
We review BiMark~\citep{feng2025bimark} as a representative unbiased multilayer watermarking framework. At each token position, $x_{-h}$ seeds the selection of a single message bit position $p \in \{1,\ldots,k\}$. BiMark embeds the selected bit $m_p$ across $\ell$ layers by constructing $\ell$ independent vocabulary bipartitions $\{(\mathcal{V}_0^i,\mathcal{V}_1^i)\}_{i=1}^{\ell}$. For layer $i$, the reweighting direction is computed as $e^i = m_p \oplus b^i$, where $b^i \sim \operatorname{Bernoulli}(0.5)$ is a pseudorandom mask. Thus, $e^i$ is a fair coin flip regardless of the embedded message bit.

Let $\theta_i$ denote the reweighting configuration of layer $i$, including the bipartition $(\mathcal{V}_0^i,\mathcal{V}_1^i)$ and the bias parameters. Starting from $P_t^{(0)} = P_M(\cdot \mid x_{<t})$, BiMark applies multilayer reweighting as $P_t^{(i)} = R_{\theta_i,e^i}(P_t^{(i-1)})$ for $i=1,\ldots,\ell$, and samples the next token from $P_{M,w}(\cdot \mid x_{<t})=P_t^{(\ell)}$.

The bit-flip reweighting preserves the input distribution in expectation. Consider a token $x \in \mathcal{V}_0^i$ and let $\delta_0$ be the bias factor applied to this partition. Depending on the reweighting direction $e^i$, the token probability $P(x)$ is scaled by either $(1+\delta_0)$ or $(1-\delta_0)$. Since $e^i$ is a fair coin flip,
\begin{equation}
\mathbb{E}_{e^i}\!\left[R_{\theta_i,e^i}(P)(x)\right]
= \frac{1}{2}(1+\delta_0)P(x)
+ \frac{1}{2}(1-\delta_0)P(x)
= P(x).
\end{equation}
The same argument holds for tokens in $\mathcal{V}_1^i$ with bias factor $\delta_1$. Hence, $\mathbb{E}_{e^i}[R_{\theta_i,e^i}(P)] = P$. Therefore, applying the layers sequentially also preserves the original distribution in expectation.

\textbf{Message extraction.}
BiMark extracts the message using a voting matrix $\mathbf{M}\in\mathbb{N}_0^{k\times 2}$. For each token $x_t$, the extractor reconstructs the assigned bit position $p$ and the mask $b^i$ from $x_{-h}$. Each layer provides one vote: the reweighting direction is estimated as $\hat{e}^i=\indicator\{x_t\in\mathcal{V}_1^i\}$, the bit estimate is recovered as $\hat{m}^i=\hat{e}^i\oplus b^i$, and the vote is updated as $\mathbf{M}[p,\hat{m}^i]\gets \mathbf{M}[p,\hat{m}^i]+1$. After processing all tokens, each message bit is extracted by majority voting as $\hat{m}_p=\arg\max_{b\in\{0,1\}}\mathbf{M}[p,b]$.

We evaluate extraction performance using bit accuracy and match rate. Bit accuracy is the fraction of embedded bits $c_p$ correctly estimated from the accumulated votes before any ECC decoding. Match rate is the proportion of samples for which the entire message
is correctly recovered, i.e., $\hat{\mathbf{m}}=\mathbf{m}$. In multi-bit watermarking, match rate is the more critical metric because the embedded message is useful only when all bits are recovered correctly.

\section{Proposed method}
\label{sec:proposed_method}

\subsection{Overview}
\label{sec:method_overview}

\textsc{WeaveMark} improves the multi-bit watermarking trade-off through two main components. First, payload spreading improves payload capacity by spreading the embedded bit sequence over generated tokens and multiple layers. It combines \emph{multi-bit-per-token embedding}, which increases token-level coverage for each embedded bit, with \emph{layer shuffling}, which balances layer reliability across bit positions. Second, \emph{soft-decision ECC} decoding improves extraction accuracy by using vote margins instead of hard majority decisions.

These components are complementary: ECC adds redundancy for message recovery but expands the number of bits that must be embedded, and payload spreading counteracts this overhead by assigning more token-level observations to each embedded bit. Together they improve payload capacity and extraction accuracy while retaining the text-quality benefit of unbiased multilayer reweighting. The complete embedding and extraction algorithms of \textsc{WeaveMark} are provided in Appendix~\ref{app:algorithm}.  

\subsection{Payload spreading}
\label{sec:payload_spreading}

\textbf{Motivation.}
In one-bit-per-token multilayer watermarking, each token assigns a bundle of $\ell$ layer-wise votes to a single bit position. When there are few tokens or many message bits, this all-at-once assignment can make the vote distribution uneven: some bits receive many vote bundles, while others receive few or none. A more balanced strategy is to divide the votes from each token across multiple bit positions.

\textbf{Multi-bit-per-token embedding.}
In \textsc{WeaveMark}, the $k$-bit message $\mathbf{m}$ is first encoded into an $n$-bit ECC codeword $\mathbf{c}=E(\mathbf{m})$. The watermarking procedure then spreads and embeds the codeword bits across generated tokens.

Instead of assigning each token to a single bit, \textsc{WeaveMark} embeds $\kappa$ bits per token. At each token position, the preceding context $x_{-h}$ seeds a pseudorandom function that selects $\kappa$ bit positions $\mathcal{P}=\{p_1,\ldots,p_\kappa\}\subseteq\{1,\ldots,n\}$. The $\ell$ layers are assigned to these positions as evenly as possible, so each selected bit receives either $\lfloor \ell/\kappa \rfloor$ or $\lfloor \ell/\kappa \rfloor+1$ layers. For a layer $i$ assigned to bit $c_p$, \textsc{WeaveMark} computes the reweighting direction as $e^i=c_p\oplus b^i$, and applies the reweighting function $R_{\theta_i,e^i}$ as in Section~\ref{sec:preliminary}.

Under uniform selection, each bit is assigned to approximately $\kappa T/n$ token positions instead of $T/n$, so while the layer budget per token remains $\ell$, each bit draws token-level statistics from $\kappa$ times more positions and vote accumulation across bits becomes more even.

\textbf{Layer shuffling.}
Although the bit positions $\mathcal{P}$ are selected to spread token-level statistics across the embedded bits, layer assignment can still introduce imbalance because the layers have different extraction reliability. 
Because multilayer reweighting is applied sequentially, earlier layers leave clearer vote signals, whereas later layers are perturbed by the preceding reweighting steps and their votes become less reliable. In our layer-wise analysis using LLaMA-3-8B on C4 RealNewsLike, extraction accuracy decreases from 72.20\% at the first layer to 56.98\% at the tenth layer. Thus, fixed layer assignment can repeatedly assign more reliable layers to some bit positions than to others.

To reduce this layer-wise imbalance, \textsc{WeaveMark} randomly permutes the layer indices using a pseudorandom function seeded by the preceding context $x_{-h}$ before assigning layers to the selected bit positions. This prevents fixed bit positions from repeatedly receiving the same high- or low-reliability layers, making the effective embedding strength more balanced.

\begin{figure}[t]
    \centering
    \includegraphics[width=\textwidth]{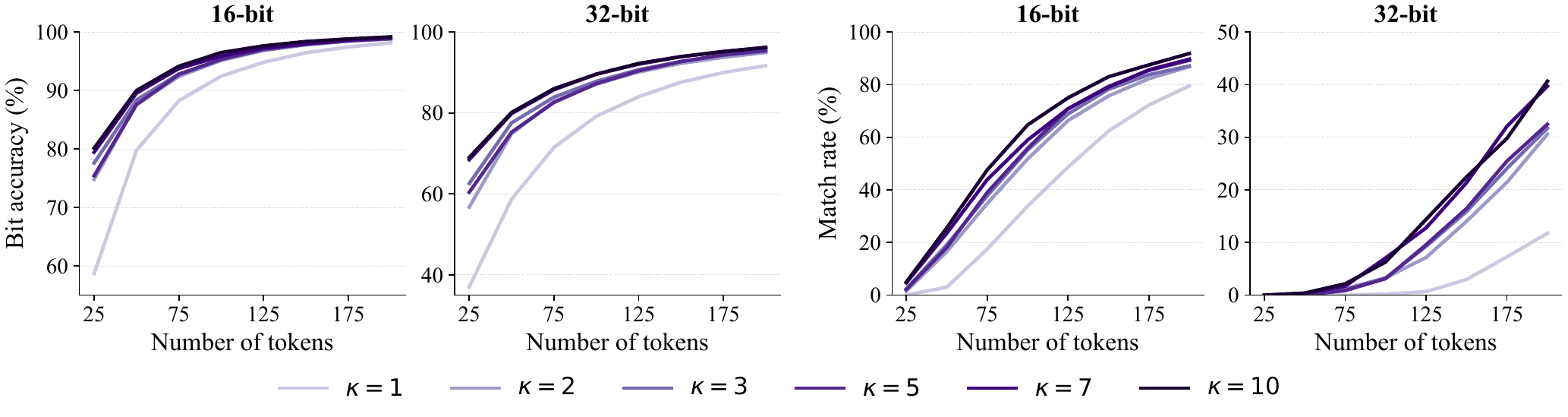}
    \vspace{-15pt}
    \caption{Effect of the number of bits embedded in each token, $\kappa$, in payload spreading. Increasing $\kappa$ spreads each bit over more token positions, improving both bit accuracy and match rate.}
    \vspace{-7pt}
    \label{fig:bpt_ablation}
\end{figure}

\textbf{Effect of payload spreading.}
Figure~\ref{fig:bpt_ablation} shows the effect of increasing the number of bits per token $\kappa$ with layer shuffling and a fixed layer budget of $\ell=10$, using Mistral-7B on C4 RealNewsLike.
Increasing $\kappa$ improves both bit accuracy and match rate, with larger gains when token-level observations are scarce. For 16-bit messages at 50 tokens, match rate improves from 3.01\% ($\kappa=1$) to 25.56\% ($\kappa=10$); for 32-bit messages at 200 tokens, it improves from 11.76\% to 40.64\%.

\begin{figure}[t]
    \centering
    \includegraphics[width=\textwidth]{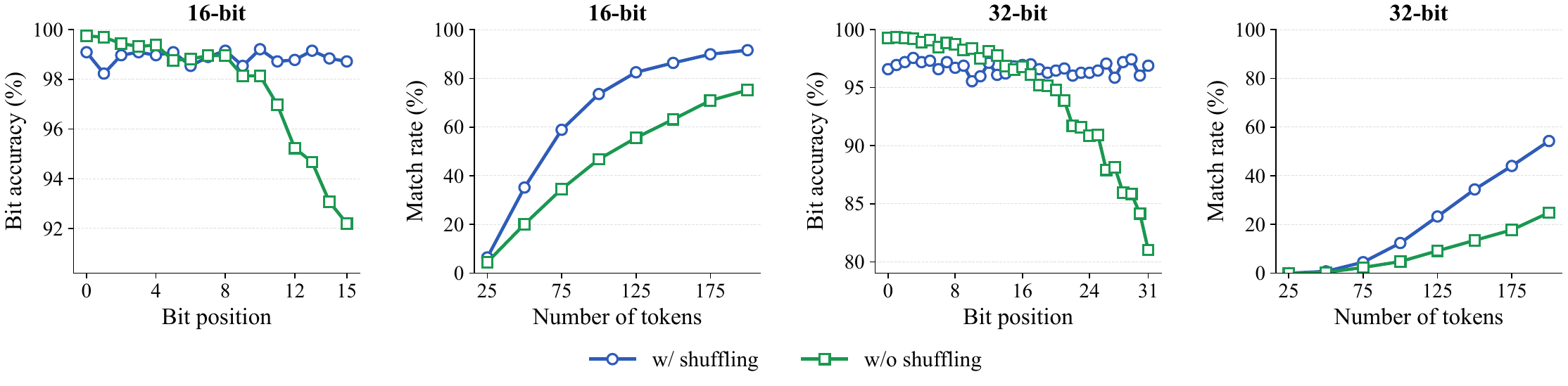}
    \vspace{-15pt}
    \caption{Effect of layer shuffling in payload spreading. Bit-position accuracy is measured at 200 tokens. Layer shuffling balances layer reliability across bit positions, improving match rate.}
    \vspace{-7pt}
    \label{fig:shuffling_layerwise}
\end{figure}

Figure~\ref{fig:shuffling_layerwise} isolates the effect of layer shuffling.
Here, bit accuracy is measured separately for each bit position.
Without shuffling, the original bit-index order consistently places higher-index bits in less reliable later layers, worsening bit-accuracy imbalance for longer messages.
Layer shuffling balances bit accuracy across positions, improving match rate from 46.8\% to 73.6\% for 16 bits and 100 tokens, and from 24.9\% to 54.3\% for 32 bits and 200 tokens.

\textbf{Unbiased property.}
Payload spreading and layer shuffling preserve the unbiased reweighting property. Let $\tilde{p}^{\,i}\in\{1,\ldots,n\}$ denote the bit position assigned to layer $i$ after payload spreading and layer shuffling, so that the reweighting direction at layer $i$ is $e^i=c_{\tilde{p}^{\,i}}\oplus b^i$. For any fixed bit selection and layer assignment, $c_{\tilde{p}^{\,i}}$ is simply a fixed binary value. Since $b^i$ is a fair coin flip, $e^i=c_{\tilde{p}^{\,i}}\oplus b^i$ is also a fair coin flip. Therefore, by the unbiasedness argument in Section~\ref{sec:preliminary}, each layer preserves its input distribution in expectation:
\begin{equation}
\mathbb{E}_{b^i}\!\left[
R_{\theta_i,c_{\tilde{p}^{\,i}}\oplus b^i}(P)
\right]
=
P.
\label{eq:unbiased_spreading}
\end{equation}
Consequently, the proposed sequential multilayer reweighting preserves the original token distribution in expectation, regardless of the bit-to-layer assignment.
\subsection{Soft-decision ECC decoding}
\label{sec:soft_ecc}

\textbf{Motivation.}
Multi-bit watermarking is evaluated by exact message recovery, so even a few bit errors can sharply reduce match rate. ECC improves recovery by adding redundancy, but at the cost of additional payload bits. Therefore, the decoder should fully exploit the extracted statistics. However, applying ECC after hard-decision majority voting discards reliability information: votes $(100,99)$ and $(100,0)$ yield the same hard decision despite very different confidence levels.

\textbf{Soft reliability extraction and decoding.}
Instead of making a hard decision at each bit position, \textsc{WeaveMark} extracts a soft reliability score from the vote margin. Let $v_0[p]$ and $v_1[p]$ be the votes for 0 and 1 at bit position $p$, and define the vote margin $s[p]=v_1[p]-v_0[p]$; its sign indicates the likely bit value and its magnitude the reliability. Given the soft vote vector $\mathbf{s}=(s[1],\ldots,s[n])$, \textsc{WeaveMark} performs vote-margin-based soft maximum-likelihood  decoding. For a linear code with codeword set $\mathcal{C}$, the decoder selects the codeword most consistent with the soft observations:
\vspace{-5pt}
\begin{equation}
\hat{\mathbf{c}}
=
\arg\max_{\mathbf{c}\in\mathcal{C}}
\sum_{p=1}^{n} s[p]\,(2c[p]-1).
\end{equation}
This rule chooses the codeword with the largest correlation to the soft vote vector, and can be implemented efficiently by enumerating $\mathcal{C}$. Longer payloads are covered by concatenating more blocks of the same short code, so the enumeration cost per block is unchanged and total decoding cost grows linearly rather than exponentially with the payload length. Appendix~\ref{app:voting} compares alternative soft metrics; the vote margin performs best, so we adopt it.

\begin{figure}[t]
    \centering
    \includegraphics[width=\textwidth]{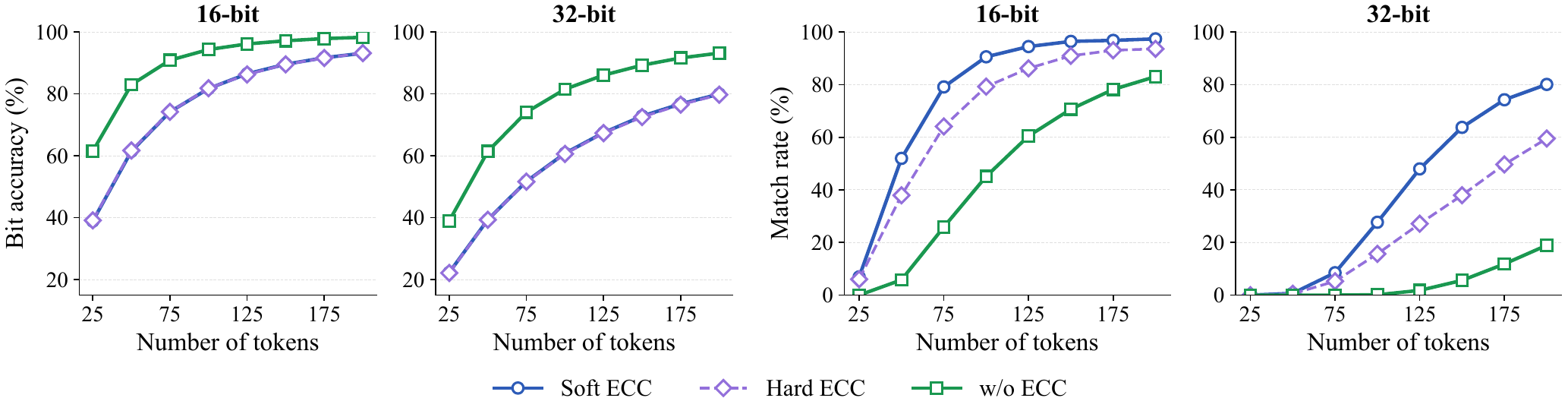}
    \vspace{-15pt}
    \caption{Effect of soft-decision ECC with one-bit-per-token embedding. Soft-decision decoding uses vote margins as reliability information, improving match rate over hard ECC and no ECC.}
    \vspace{-9pt}
    \label{fig:ecc_ablation}
\end{figure}

\textbf{Effect of soft-decision ECC.}
Figure~\ref{fig:ecc_ablation} shows the effect of adding soft-decision ECC to one-bit-per-token embedding on OpenGen. We use rate-1/2 block codes: $(24,12)$ extended Golay~\citep{golay1949notes} for 12-bit messages, $(32,16)$ Reed-Muller~\citep{reed1954class,muller1954application} for 16-bit messages, and two independent blocks for 24- and 32-bit messages.
ECC increases the number of embedded bits and can reduce bit accuracy, but its redundancy substantially improves match rate, especially for longer messages. For 32-bit messages at 200 tokens, match rate improves from 18.94\% to 80.07\%.

\subsection{Component ablation}
\label{sec:component_ablation}
Table~\ref{tab:component_ablation} summarizes the cumulative effect of each component.
Starting from BiMark, multi-bit-per-token embedding modestly improves match rate by spreading token-level statistics across bit positions.
Layer shuffling then improves exact recovery by balancing layer reliability across bit positions.
Finally, soft-decision ECC lowers bit accuracy by adding parity bits, but its error-correction capability more than compensates for this overhead and further boosts match rate.
Together, these components improve match rate from 20.78\% to 89.84\% for 32-bit messages with 200 tokens.
With the same soft ECC, \textsc{WeaveMark} outperforms BiMark by 9.25 percentage points in match rate, demonstrating the independent effect of payload spreading. The components are complementary: multi-bit embedding and layer shuffling balance the reliability of observations across the expanded ECC codeword, enabling more effective soft decoding under a fixed token budget.

\begin{table}[!ht]
\centering
\caption{Component ablation for 32-bit messages with 200 generated tokens.}
\label{tab:component_ablation}
\vspace{-5pt}
\tablefootnotesize
\setlength{\tabcolsep}{5pt}
\renewcommand{\arraystretch}{1.0}
\begin{tabular}{lccccc}
\toprule
Method & Multi-bit & Shuffling & Soft ECC & Bit accuracy (\%) & Match rate (\%) \\
\midrule
BiMark (Baseline) & -- & -- & -- & 93.21 & 20.78 \\
+ Multi-bit embedding & \checkmark & -- & -- & 94.52 & 24.00 \\
+ Layer shuffling & \checkmark & \checkmark & -- & 96.67 & 55.88 \\
+ Soft ECC (\textsc{WeaveMark}) & \checkmark & \checkmark & \checkmark & 90.51 & \textbf{89.84} \\
BiMark + Soft ECC & -- & -- & \checkmark & 79.78 & 80.59 \\
\bottomrule
\end{tabular}
\end{table}
\section{Experiments}
\label{sec:experiments}

\subsection{Experimental setup}

\textbf{Models and datasets.}
We use LLaMA-3-8B~\citep{dubey2024llama3} as the primary model for extraction 
and robustness experiments, and Gemma-2-9B~\citep{team2024gemma2} for perplexity 
evaluation. We evaluate mainly on C4 RealNewsLike~\citep{raffel2020t5}, 
and additionally use Mistral-7B~\citep{jiang2023mistral} and 
OpenGen~\citep{krishna2023paraphrasing} for ablation experiments.

\textbf{Baselines.}
We compare \textsc{WeaveMark} against BiMark~\citep{feng2025bimark}, MPAC~\citep{yoo2024mpac}, and RSBH~\citep{qu2025provably}. For MPAC and RSBH, we evaluate $\delta\in\{2.0,3.0\}$ to show the trade-off between extraction accuracy and text quality.
For BiMark and \textsc{WeaveMark}, we use $\delta=1.0$, $\ell=10$, and $h=2$ (Appendix~\ref{app:window}), with $\kappa=10$ for \textsc{WeaveMark}; BiMark embeds one message bit per token. For MPAC and RSBH, we use their recommended $h=1$, with $r=4$ and $\gamma=0.25$ for MPAC and $\gamma=0.5$ for RSBH. We use top-$K$ sampling with $K=50$ and temperature $\tau=1.0$.

\subsection{Message extraction accuracy}
\label{sec:extraction}

We first evaluate whether \textsc{WeaveMark} improves exact message recovery on clean watermarked text and under editing attacks. The clean setting measures payload scalability, while the attack setting evaluates robustness when token-level statistics are partially corrupted.

\begin{figure}[t]
    \centering
    \includegraphics[width=0.95\textwidth]{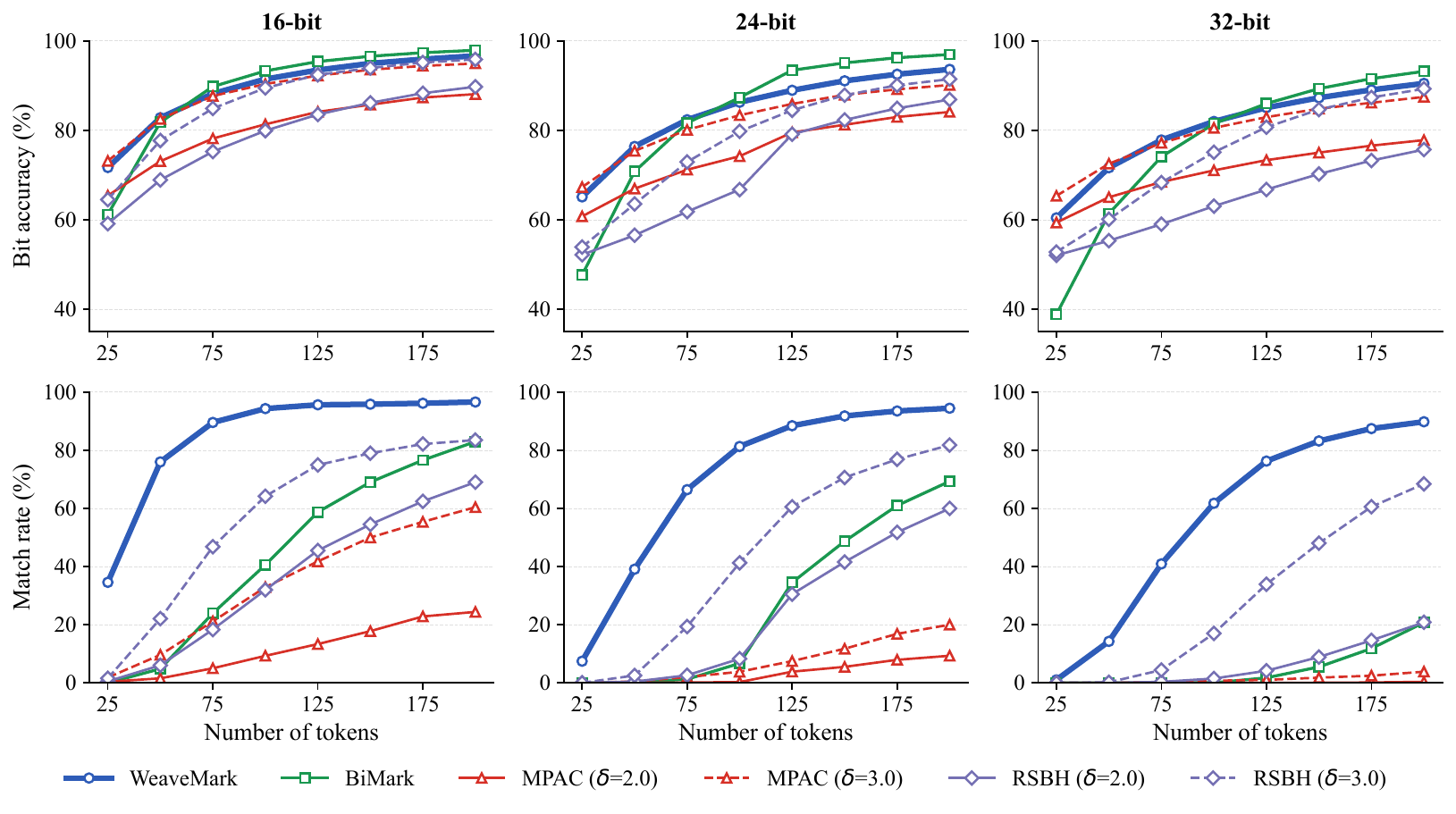}
    \caption{Clean-text message extraction accuracy. \textsc{WeaveMark} improves match rate across token lengths, enabling reliable 32-bit payload recovery beyond the range of prior methods.}
    \label{fig:extraction}
\end{figure}

\textbf{Clean-text extraction.}
Figure~\ref{fig:extraction} shows bit accuracy and match rate on clean text. \textsc{WeaveMark} consistently improves match rate over the baselines, especially for long messages and short generated texts. For short generated texts of 50 tokens, \textsc{WeaveMark} achieves 76.02\% match rate for 16-bit messages, while BiMark drops to 4.84\%. For long messages, \textsc{WeaveMark} achieves 89.84\% match rate for 32-bit messages with 200 tokens, compared with 20.78\% for BiMark. MPAC and RSBH underperform even with $\delta=3.0$, which uses stronger biased reweighting. Additional results in Appendix~\ref{app:scalability} demonstrate scalability to longer messages of up to 64 bits.

\begin{figure}[t]
    \centering
    \includegraphics[width=\textwidth]{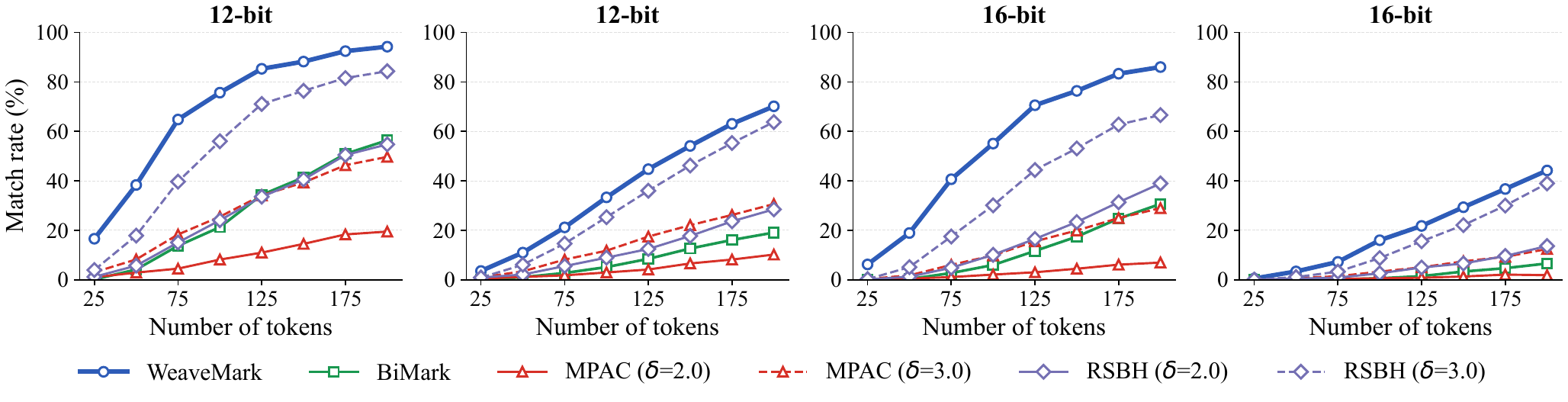}
    \vspace{-15pt}
    \caption{Robustness to synonym substitution attacks. Within each message-length group, the left and right panels use substitution ratios $\rho=0.1$ and $\rho=0.2$, respectively.}
    \label{fig:robustness}
\end{figure}

\textbf{Robustness to editing attacks.}
\looseness=-1 %
We evaluate robustness under synonym substitution attacks, where a fraction $\rho$ of tokens are replaced with synonyms using WordNet~\citep{miller1995wordnet} candidates filtered by BERT~\citep{devlin2019bert}. 
Figure~\ref{fig:robustness} shows the match rate under different substitution ratios. For 16-bit messages with 200 generated tokens at $\rho=0.1$, \textsc{WeaveMark} achieves 86.02\% match rate, compared with 30.67\% for BiMark. At $\rho=0.2$, \textsc{WeaveMark} achieves 44.20\%, while BiMark drops to 6.59\%.
Although RSBH with $\delta=3.0$ shows competitive robustness in some settings, this relies on stronger biased reweighting and leads to text-quality degradation, as shown in Section~\ref{sec:quality}. 
Appendix~\ref{app:robustness} provides additional robustness results for various message lengths and substitution ratios, as well as insertion, deletion, mixed edits, truncation, copy-paste, and rewriting attacks.

\subsection{Text quality}
\label{sec:quality}

Table~\ref{tab:quality} reports text quality on two downstream tasks: summarization with BART-large-CNN~\citep{lewis2020bart} on CNN/DailyMail~\citep{hermann2015teaching}, measured by BERTScore~\citep{zhang2020bertscore} (RoBERTa-large~\citep{liu2019roberta}) and ROUGE~\citep{lin2004rouge}; and translation with mBART-large-50~\citep{tang2020multilingual} on WMT16 en$\to$ro~\citep{bojar2016findings}, measured by BERTScore (XLM-RoBERTa-large~\citep{conneau2020xlmr}) and BLEU~\citep{papineni2002bleu}.

\textsc{WeaveMark} achieves text quality comparable to unwatermarked text on most metrics.
Unbiased reweighting preserves the expected token distribution, as in Eq.~\eqref{eq:unbiased_spreading}, but allows individual outputs to differ.
BLEU, which measures exact $n$-gram matches to reference texts, is slightly lower than for unwatermarked text but comparable to the unbiased baseline BiMark. 
The comparable BERTScore and ROUGE scores support overall text quality preservation in terms of semantic fidelity and content coverage.
In contrast, MPAC and RSBH show degradation across all metrics, worsening with stronger biases.
Appendix~\ref{app:perplexity} provides perplexity analysis.

\begin{table}[htbp]
\centering
\caption{Text quality on downstream tasks. Higher values indicate better quality.}
\label{tab:quality}
\tablefootnotesize
\setlength{\tabcolsep}{8pt}
\renewcommand{\arraystretch}{1.2}
\begin{tabular}{lcccccc}
\toprule
& \multicolumn{4}{c}{Summarization} & \multicolumn{2}{c}{Translation} \\
\cmidrule(lr){2-5} \cmidrule(lr){6-7}
Method & BERTScore & R-1 & R-2 & R-L & BERTScore & BLEU \\
\midrule
No watermark & 30.54 & 41.15 & 19.17 & 28.63 & 94.95 & 25.24 \\
\midrule
\textsc{WeaveMark} & 30.48 & 40.89 & 18.22 & 28.11 & 94.52 & 23.15 \\
BiMark & 30.27 & 40.70 & 18.28 & 28.00 & 94.56 & 23.22 \\
MPAC ($\delta=2.0$) & 29.81 & 39.75 & 16.87 & 26.97 & 94.39 & 21.47 \\
MPAC ($\delta=3.0$) & 26.53 & 36.61 & 12.87 & 23.60 & 93.50 & 16.78 \\
RSBH ($\delta=2.0$) & 29.58 & 39.74 & 16.62 & 26.91 & 94.37 & 21.02 \\
RSBH ($\delta=3.0$) & 27.45 & 37.69 & 14.07 & 24.61 & 93.75 & 16.94 \\
\bottomrule
\end{tabular}
\end{table}

\section{Zero-bit detection}
\label{sec:zerobit}
Multi-bit watermarking biases token sampling toward a message-dependent vocabulary partition, which we call the target partition. For zero-bit detection, the detector must count how often generated tokens fall into this target partition. However, in multi-bit watermarking, the embedded message is unknown during detection, so the target partition cannot be reconstructed directly.

Existing schemes such as BiMark and MPAC address this by treating the most-voted partition as the target and then computing a $z$-score. However, this max-over-partitions operation inflates the statistic even for non-watermarked text, raising the threshold required to control the false-positive rate and reducing detection power. This effect becomes more severe as the message length increases.

To avoid this issue, \textsc{WeaveMark} reserves a few dedicated zero-bit layers that do not encode the message. Their target partitions are determined by the context, as in \citet{kirchenbauer2023watermark}, rather than by the message bit. Therefore, the detector can directly reconstruct the target partition. These layers use the same unbiased reweighting mechanism, preserving the expected token distribution.

Detection applies the standard $z$-test with $\gamma=0.5$ to the zero-bit layers. The known target partition eliminates the need for a max-over-partition operation, keeping the detection threshold low and stable at a fixed FPR.
With $\ell$ total layers, increasing the number of zero-bit layers $\ell_z$ strengthens presence detection but leaves fewer layers $\ell-\ell_z$ for multi-bit embedding.

\begin{figure}[t]
    \centering
    \includegraphics[width=\textwidth]{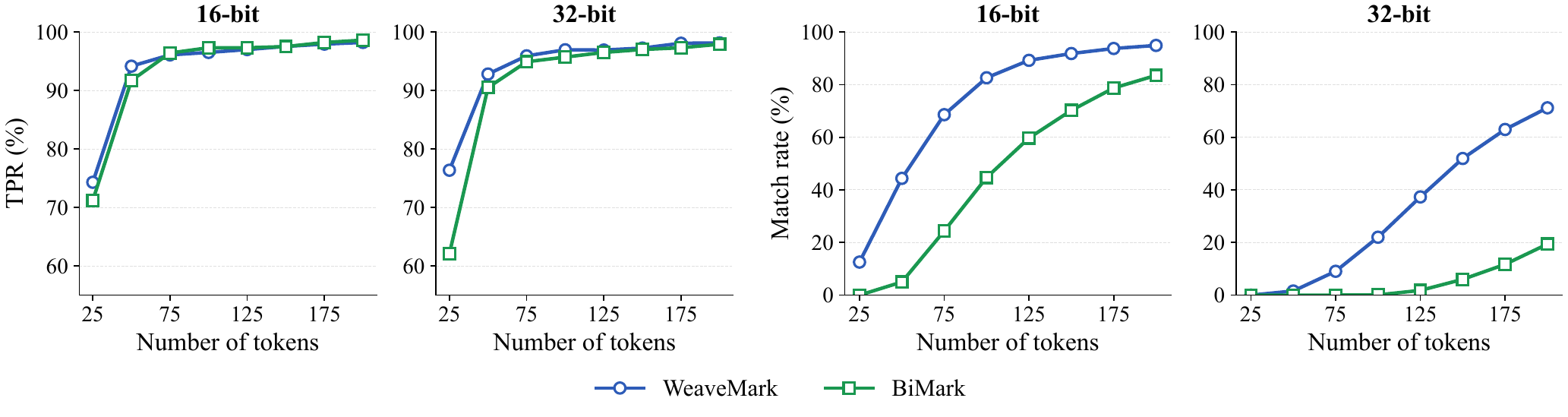}
    \vspace{-15pt}
        \caption{Zero-bit detection and multi-bit message recovery at FPR=1\% with $\ell=10$ layers.
\textsc{WeaveMark} uses $\ell_z=2$ dedicated detection layers, while BiMark uses message-bit votes across all ten.}
    \label{fig:zerobit_multibit}
\end{figure}

\paragraph{Multi-bit setting.}
Figure~\ref{fig:zerobit_multibit} compares watermark presence detection and message recovery at FPR=1\%, with both methods using $\ell=10$ total layers. \textsc{WeaveMark} dedicates $\ell_z=2$ layers to presence detection, whereas BiMark uses a max-over-partition statistic from message-bit votes across all ten layers. Despite using only two layers for detection, \textsc{WeaveMark} achieves TPR comparable to BiMark, with both attaining high TPR beyond 50 tokens. With comparable zero-bit detection performance, \textsc{WeaveMark} retains a substantial gain in multi-bit message recovery: for 32-bit messages with 200 tokens, it achieves a match rate of 71.17\%, compared with 19.40\% for BiMark.

\noindent
\begin{minipage}[t]{0.61\textwidth}\vspace{0pt}
Table~\ref{tab:zerobit_zthr} reports the $z$-score thresholds required for FPR=1\%. BiMark requires a higher threshold for 32-bit than for 16-bit messages because its max-over-partition statistic becomes more inflated on non-watermarked text as message length increases.
In contrast, \textsc{WeaveMark}'s threshold remains near 2.3 because its dedicated zero-bit layers are independent of the multi-bit codeword structure.
Thus, larger payloads do not require a comparable increase in its detection threshold.

\end{minipage}\hfill
\begin{minipage}[t]{0.37\textwidth}\vspace{0pt}\centering
\vspace{-\abovecaptionskip}
\captionof{table}{$z$-score threshold at FPR=1\% averaged over 25 to 200 tokens.}
\label{tab:zerobit_zthr}
\vspace{0.75em}
\small
\setlength{\tabcolsep}{6pt}
\renewcommand{\arraystretch}{1.15}
\begin{tabular}{lcc}
\toprule
Method & 16-bit & 32-bit \\
\midrule
\textsc{WeaveMark} & \textbf{2.24} & \textbf{2.33} \\
BiMark & 4.54 & 5.55 \\
\bottomrule
\end{tabular}
\end{minipage}

\textbf{Pure zero-bit setting.}
Table~\ref{tab:zerobit_compare} reports TPR at FPR=1\% against KGW soft red-green~\citep{kirchenbauer2023watermark}, SynthID~\citep{dathathri2024synthid}, DiPmark~\citep{wu2024dipmark}, and EXP-edit~\citep{kuditipudi2024robust}.
Here, \textsc{WeaveMark} uses all layers for zero-bit detection $\ell_z=10$.
Despite supporting flexible layer allocation between zero-bit and multi-bit embedding, it performs on par with the strongest schemes designed exclusively for zero-bit detection.

\begin{table}[h]
\centering
\caption{Pure zero-bit detection TPR (\%) at FPR=1\%. Columns give the number of generated tokens.}
\label{tab:zerobit_compare}
\small
\tablefootnotesize
\setlength{\tabcolsep}{6pt}
\renewcommand{\arraystretch}{1.2}
\begin{tabular*}{\textwidth}{@{\extracolsep{\fill}}lccccccc@{}}
\toprule
Method & 50 & 75 & 100 & 125 & 150 & 175 & 200 \\
\midrule
\textsc{WeaveMark} & 97.94 & 98.53 & 98.67 & 98.45 & 98.81 & 98.99 & 98.95 \\
EXP-edit           & 99.40 & 99.60 & 99.63 & 99.53 & 99.77 & 99.80 & 99.93 \\
SynthID            & 98.09 & 98.41 & 98.61 & 98.68 & 98.92 & 99.08 & 99.26 \\
KGW ($\delta=2.0$) & 92.98 & 96.74 & 97.49 & 97.67 & 98.10 & 98.61 & 99.13 \\
DiPmark ($\alpha=0.45$) & 46.43 & 72.21 & 78.29 & 85.61 & 88.46 & 89.70 & 91.60 \\
\bottomrule
\end{tabular*}
\end{table}
\section{Conclusion}
We presented \textsc{WeaveMark}, a robust and scalable multi-bit LLM watermarking scheme using coded payload spreading. \textsc{WeaveMark} improves payload capacity through multi-bit-per-token spreading, improves extraction accuracy through soft-decision ECC, and preserves text quality through unbiased multilayer reweighting.
Experiments show that \textsc{WeaveMark} substantially improves match rate for long messages, short texts, and editing attacks without degrading text quality. It improves clean-text match rate from 20.8\% to 89.8\% and robustness under synonym substitution from 30.7\% to 86.0\%.
Flexible layer allocation enables strong zero-bit detection alongside accurate multi-bit recovery, while matching leading zero-bit schemes in the pure zero-bit setting.
These results demonstrate that \textsc{WeaveMark} supports source tracing and watermark detection within a unified framework.

\subsection*{Reproducibility statement}
The complete embedding and extraction procedures of \textsc{WeaveMark} are given as Algorithms~\ref{alg:embed} and~\ref{alg:extract} in Appendix~\ref{app:algorithm}, and the dedicated zero-bit layers are specified in Section~\ref{sec:zerobit}. Section~\ref{sec:experiments} lists the models, datasets, baselines, sampling configuration, and all hyperparameters; Section~\ref{sec:soft_ecc} specifies the error-correcting code used for each message length; and Section~\ref{sec:quality} describes the text-quality evaluation. The attack configurations and the settings of the extended comparison are given in Section~\ref{sec:extraction} and Appendices~\ref{app:extended}--\ref{app:robustness}. All models and datasets used are publicly available. Our implementation is available at the link given in the footnote on the first page.

\bibliography{references}
\bibliographystyle{iclr2027_conference}

\newpage
\appendix
\section{Algorithm details}
\label{app:algorithm}

Algorithm~\ref{alg:embed} describes the multi-bit message embedding procedure used for source tracing, and Algorithm~\ref{alg:extract} describes the corresponding message extraction procedure. The dedicated zero-bit layers used for watermark presence detection are independent of the message bits and are described separately in Section~\ref{sec:zerobit}.

\begin{algorithm}[h]
\caption{Message Embedding}
\label{alg:embed}
\begin{flushleft}
\textbf{Input:} \\
\quad a language model $M$ \\
\quad an initial prompt $x_1,\ldots,x_L$ with $L\ge h$ \\
\quad a sequence of balanced vocabulary bipartitions $\{(\mathcal{V}_0^i,\mathcal{V}_1^i)\}_{i=1}^{\ell}$ \\
\quad reweighting configurations $\{\theta_i\}_{i=1}^{\ell}$, where $\theta_i$ includes $(\mathcal{V}_0^i,\mathcal{V}_1^i)$ and bias parameters \\
\quad a message $\mathbf{m}\in\{0,1\}^k$ \\
\quad an ECC encoder $E:\{0,1\}^k\to\{0,1\}^n$ \\
\quad window size $h$ and bits per token $\kappa$, where $1\le \kappa\le \ell$ \\
\quad a bit-flip unbiased reweighting function $R$ \\
\quad pseudorandom functions $\mathrm{PRF}_b$, $\mathrm{PRF}_p$, and $\mathrm{PRF}_s$
\end{flushleft}
\hrule
\vspace{0.5em}
\begin{flushleft}
1. Encode the message into a codeword:
$\mathbf{c}\leftarrow E(\mathbf{m})$, where $\mathbf{c}\in\{0,1\}^n$. \\[0.5em]

2. Initialize the set of used contexts as $\mathcal{S}\leftarrow\emptyset$. \\[0.5em]

\textbf{for} $t=L+1,L+2,\ldots$ \textbf{do}: \\[0.3em]

\quad 3. Compute the original next-token distribution
$P_t^{(0)}\leftarrow P_M(\cdot\mid x_{<t})$. \\[0.5em]

\quad 4. Let $x_{-h}^{(t)}=x_{t-h:t-1}$ be the preceding $h$-token context. \\
\quad \quad \textbf{if} $x_{-h}^{(t)}\in\mathcal{S}$ \textbf{then} \\
\quad \quad \quad sample $x_t\sim P_t^{(0)}$ and continue. \\
\quad \quad \textbf{else} \\
\quad \quad \quad update $\mathcal{S}\leftarrow\mathcal{S}\cup\{x_{-h}^{(t)}\}$ and compute \\
\quad \quad \quad $(\mathrm{seed}_b,\mathrm{seed}_p,\mathrm{seed}_s)
\leftarrow
(\mathrm{PRF}_b(x_{-h}^{(t)}),\mathrm{PRF}_p(x_{-h}^{(t)}),\mathrm{PRF}_s(x_{-h}^{(t)}))$. \\
\quad \quad \textbf{end if} \\[0.5em]

\quad 5. Generate pseudorandom mask bits $b^1,\ldots,b^{\ell}$ from $\mathrm{seed}_b$. \\
\quad \quad Select $\kappa$ distinct bit positions
$\mathcal{P}=\{p_1,\ldots,p_\kappa\}\subseteq\{1,\ldots,n\}$ using $\mathrm{seed}_p$. \\[0.5em]

\quad 6. Construct a balanced layer-assignment vector
$\tilde{\mathbf{p}}=(\tilde{p}^{\,1},\ldots,\tilde{p}^{\,\ell})$. \\
\quad \quad Let $q=\lfloor \ell/\kappa\rfloor$ and $r=\ell\bmod \kappa$. \\
\quad \quad Include $q+1$ copies of each of $p_1,\ldots,p_r$ in $\tilde{\mathbf{p}}$. \\
\quad \quad Include $q$ copies of each of $p_{r+1},\ldots,p_\kappa$ in $\tilde{\mathbf{p}}$. \\
\quad \quad Shuffle the entries of $\tilde{\mathbf{p}}$ using $\mathrm{seed}_s$. \\[0.5em]

\quad 7. \textbf{for} $i=1,2,\ldots,\ell$ \textbf{do}: \\
\quad \quad \quad Compute the reweighting direction
$e^i\leftarrow c_{\tilde{p}^{\,i}}\oplus b^i$. \\
\quad \quad \quad Apply the $i$-th reweighting layer:
$P_t^{(i)}\leftarrow R_{\theta_i,e^i}(P_t^{(i-1)})$. \\
\quad \quad \textbf{end for} \\[0.5em]

\quad 8. Sample the next token $x_t\sim P_t^{(\ell)}$. \\[0.3em]

\textbf{end for}
\end{flushleft}
\end{algorithm}

\begin{algorithm}[H]
\caption{Message Extraction}
\label{alg:extract}
\begin{flushleft}
\textbf{Input:} \\
\quad a token sequence $x_1,x_2,\ldots,x_T$ \\
\quad a sequence of balanced vocabulary bipartitions $\{(\mathcal{V}_0^i,\mathcal{V}_1^i)\}_{i=1}^{\ell}$ \\
\quad codeword length $n$ \\
\quad window size $h$ and bits per token $\kappa$, where $1\le \kappa\le \ell$ \\
\quad an ECC decoder $D:\mathbb{R}^n\to\{0,1\}^k$ based on soft-decision decoding \\
\quad pseudorandom functions $\mathrm{PRF}_b$, $\mathrm{PRF}_p$, and $\mathrm{PRF}_s$
\end{flushleft}
\hrule
\vspace{0.5em}
\begin{flushleft}
1. Initialize vote counts $v_0[p]\leftarrow 0$ and $v_1[p]\leftarrow 0$ for all $p\in\{1,\ldots,n\}$. \\[0.5em]

2. Initialize the set of used contexts as $\mathcal{S}\leftarrow\emptyset$. \\[0.5em]

\textbf{for} $t=h+1,h+2,\ldots,T$ \textbf{do}: \\[0.3em]

\quad 3. Let $x_{-h}^{(t)}=x_{t-h:t-1}$ be the preceding $h$-token context. \\
\quad \quad \textbf{if} $x_{-h}^{(t)}\in\mathcal{S}$ \textbf{then} \\
\quad \quad \quad skip the current token and continue. \\
\quad \quad \textbf{else} \\
\quad \quad \quad update $\mathcal{S}\leftarrow\mathcal{S}\cup\{x_{-h}^{(t)}\}$ and compute \\
\quad \quad \quad $(\mathrm{seed}_b,\mathrm{seed}_p,\mathrm{seed}_s)
\leftarrow
(\mathrm{PRF}_b(x_{-h}^{(t)}),\mathrm{PRF}_p(x_{-h}^{(t)}),\mathrm{PRF}_s(x_{-h}^{(t)}))$. \\
\quad \quad \textbf{end if} \\[0.5em]

\quad 4. Generate pseudorandom mask bits $b^1,\ldots,b^{\ell}$ from $\mathrm{seed}_b$. \\
\quad \quad Select $\kappa$ distinct bit positions
$\mathcal{P}=\{p_1,\ldots,p_\kappa\}\subseteq\{1,\ldots,n\}$ using $\mathrm{seed}_p$. \\[0.5em]

\quad 5. Construct the same balanced layer-assignment vector
$\tilde{\mathbf{p}}=(\tilde{p}^{\,1},\ldots,\tilde{p}^{\,\ell})$. \\
\quad \quad Let $q=\lfloor \ell/\kappa\rfloor$ and $r=\ell\bmod \kappa$. \\
\quad \quad Include $q+1$ copies of each of $p_1,\ldots,p_r$ in $\tilde{\mathbf{p}}$. \\
\quad \quad Include $q$ copies of each of $p_{r+1},\ldots,p_\kappa$ in $\tilde{\mathbf{p}}$. \\
\quad \quad Shuffle the entries of $\tilde{\mathbf{p}}$ using $\mathrm{seed}_s$. \\[0.5em]

\quad 6. \textbf{for} $i=1,2,\ldots,\ell$ \textbf{do}: \\
\quad \quad \quad Estimate the reweighting direction:
$\hat{e}^i\leftarrow\indicator\{x_t\in\mathcal{V}_1^i\}$. \\
\quad \quad \quad Recover the corresponding codeword-bit estimate:
$\hat{c}_{\tilde{p}^{\,i}}\leftarrow \hat{e}^i\oplus b^i$. \\
\quad \quad \quad Update the vote count:
$v_{\hat{c}_{\tilde{p}^{\,i}}}[\tilde{p}^{\,i}]
\leftarrow
v_{\hat{c}_{\tilde{p}^{\,i}}}[\tilde{p}^{\,i}]+1$. \\
\quad \quad \textbf{end for} \\[0.3em]

\textbf{end for} \\[0.5em]

7. Compute the soft-vote vector $\mathbf{s}=(s[1],\ldots,s[n])$ by
$s[p]\leftarrow v_1[p]-v_0[p]$ for all $p\in\{1,\ldots,n\}$. \\[0.5em]

8. Decode the message from the soft-vote vector:
$\hat{\mathbf{m}}\leftarrow D(\mathbf{s})$. \\[0.5em]

9. Return the extracted message $\hat{\mathbf{m}}$.
\end{flushleft}
\end{algorithm}

\section{Extended comparison}
\label{app:extended}

\textbf{Additional baselines.}
In addition to the baselines of Section~\ref{sec:experiments}, we compare against two recent multi-bit schemes.
XMark~\citep{xu2026xmark} puts in the green list every vocabulary partition except the one indexed by the segment value. It intersects the green lists obtained from multiple secret keys and adds a positive bias to the tokens in this intersection; like MPAC and RSBH, it relies on biased reweighting.
StealthInk~\citep{jiang2025stealthink} zeroes the probability mass of one interval of a permutation of the vocabulary while doubling that of a symmetric interval. Like BiMark, it relies on unbiased reweighting, which preserves the expected token distribution.

\textbf{Setting.}
Unless stated otherwise, the comparisons in this appendix use Mistral-7B on OpenGen with a shared context window of $h=2$ across all methods (Section~\ref{sec:experiments} uses $h=1$ for MPAC and RSBH, as recommended in their respective papers). These additional experiments complement the main evaluation under method-specific settings.
XMark is configured with two secret keys and 2-bit segments, and StealthInk with 1-bit segments. For MPAC, RSBH, and XMark we use $\delta=2.0$; StealthInk has no bias parameter. Sampling otherwise follows Section~\ref{sec:experiments}.

\subsection{Payload scalability}
\label{app:scalability}

Section~\ref{sec:extraction} evaluates payloads up to 32 bits. Table~\ref{tab:scalability} extends the range from 12 to 64 bits, using one constituent block for 12- and 16-bit payloads and two, three, and four blocks for 32-, 48-, and 64-bit payloads.

\textsc{WeaveMark} achieves the highest match rate at every evaluated
payload length and token budget, with substantial gains for long payloads. At 32 bits and 200 tokens it recovers 89.4\% against 17.1\% for the strongest baseline, and at 64 bits and 500 tokens 89.3\% against 16.7\%. Equivalently, \textsc{WeaveMark} reaches a usable operating point with far fewer tokens than any baseline requires: at 400 tokens it recovers 92.8\% of 48-bit payloads, whereas the strongest baseline reaches only 42.8\% even with 500 tokens.
These results demonstrate that \textsc{WeaveMark} scales to long payloads of up to 64 bits while achieving substantially higher exact-recovery rates than existing methods.

\begin{table}[h]
\centering
\caption{Match rate (\%) from 12- to 64-bit payloads. Column groups give the payload length and the columns within each group give the number of generated tokens.}
\label{tab:scalability}
\tablefootnotesize
\setlength{\tabcolsep}{3pt}
\renewcommand{\arraystretch}{1.15}
\begin{tabular*}{\textwidth}{@{\extracolsep{\fill}}lccccccccccccccc@{}}
\toprule
& \multicolumn{3}{c}{12-bit} & \multicolumn{3}{c}{16-bit} & \multicolumn{3}{c}{32-bit} & \multicolumn{3}{c}{48-bit} & \multicolumn{3}{c}{64-bit} \\
\cmidrule(lr){2-4}\cmidrule(lr){5-7}\cmidrule(lr){8-10}\cmidrule(lr){11-13}\cmidrule(lr){14-16}
Method & 100 & 150 & 200 & 100 & 150 & 200 & 100 & 150 & 200 & 200 & 400 & 500 & 200 & 400 & 500 \\
\midrule
\textsc{WeaveMark} & \textbf{97.7} & \textbf{98.8} & \textbf{99.1} & \textbf{92.8} & \textbf{97.9} & \textbf{99.0} & \textbf{48.0} & \textbf{76.3} & \textbf{89.4} & \textbf{57.8} & \textbf{92.8} & \textbf{96.9} & \textbf{25.6} & \textbf{78.7} & \textbf{89.3} \\
BiMark     & 62.6 & 83.3 & 92.6 & 35.8 & 61.7 & 77.5 & 0.2 & 3.2 & 11.1 & 0.1 & 13.9 & 27.8 & 0.0 & 1.0 & 5.3 \\
RSBH       & 48.2 & 75.3 & 86.1 & 28.3 & 56.1 & 73.9 & 0.6 & 5.6 & 17.1 & 0.3 & 23.2 & 42.8 & 0.0 & 5.3 & 16.7 \\
MPAC       & 58.5 & 77.1 & 87.4 & 28.7 & 48.3 & 66.1 & 0.3 & 3.0 & 7.3  & 0.3 & 7.0  & 14.0 & 0.0 & 0.9 & 2.2 \\
XMark      & 41.7 & 63.3 & 76.8 & 18.4 & 38.0 & 54.1 & 0.1 & 1.4 & 3.4  & 0.1 & 3.0  & 8.8  & 0.0 & 0.2 & 0.7 \\
StealthInk & 23.6 & 43.5 & 57.8 & 9.3  & 19.2 & 31.0 & 0.0 & 0.2 & 0.8  & 0.0 & 0.4  & 1.5  & 0.0 & 0.0 & 0.0 \\
\bottomrule
\end{tabular*}
\end{table}

\subsection{Generalization across model families}
\label{app:models}

The results above use Mistral-7B, while Section~\ref{sec:experiments} uses LLaMA-3-8B as the primary model. To check that the gains are not specific to one architecture, we repeat the clean-text and substitution experiments on Qwen2.5-7B~\citep{yang2024qwen25} under the same setting. Table~\ref{tab:model_family} reports match rates for 16-bit messages with 200 generated tokens.

\textsc{WeaveMark} achieves the highest match rate on both models under both conditions, with a margin of 54.9--56.8 percentage points over the strongest baseline under substitution. This matches the LLaMA-3-8B result in Section~\ref{sec:extraction}, indicating that the gain is not specific to one model family.

\begin{table}[h]
\centering
\caption{Match rate (\%) across model families, for 16-bit messages with 200 generated tokens. Substitution uses a ratio of $\rho=0.1$.}
\label{tab:model_family}
\small
\setlength{\tabcolsep}{6pt}
\renewcommand{\arraystretch}{1.15}
\begin{tabular*}{\textwidth}{@{\extracolsep{\fill}}lcccc@{}}
\toprule
& \multicolumn{2}{c}{Mistral-7B} & \multicolumn{2}{c}{Qwen2.5-7B} \\
\cmidrule(lr){2-3}\cmidrule(lr){4-5}
Method & Clean & $\rho{=}0.1$ & Clean & $\rho{=}0.1$ \\
\midrule
\textsc{WeaveMark} & \textbf{99.0} & \textbf{81.8} & \textbf{99.1} & \textbf{84.3} \\
BiMark            & 77.5 & 25.0 & 83.0 & 29.4 \\
RSBH              & 73.9 & 19.6 & 65.3 & 20.0 \\
MPAC              & 66.1 & 23.8 & 68.4 & 25.1 \\
XMark             & 54.1 & 11.6 & 57.1 & 13.5 \\
StealthInk        & 31.0 & 3.7  & 27.1 & 3.8  \\
\bottomrule
\end{tabular*}
\end{table}

\newpage

\section{Additional robustness results}
\label{app:robustness}

The synonym substitution and DIPPER experiments follow the setting of Section~\ref{sec:experiments}; all other experiments in this appendix follow the extended-comparison setting in Appendix~\ref{app:extended}.

\subsection{Synonym substitution}
\label{app:substitution}

\begin{figure}[H]
    \centering
    \includegraphics[width=0.85\textwidth]{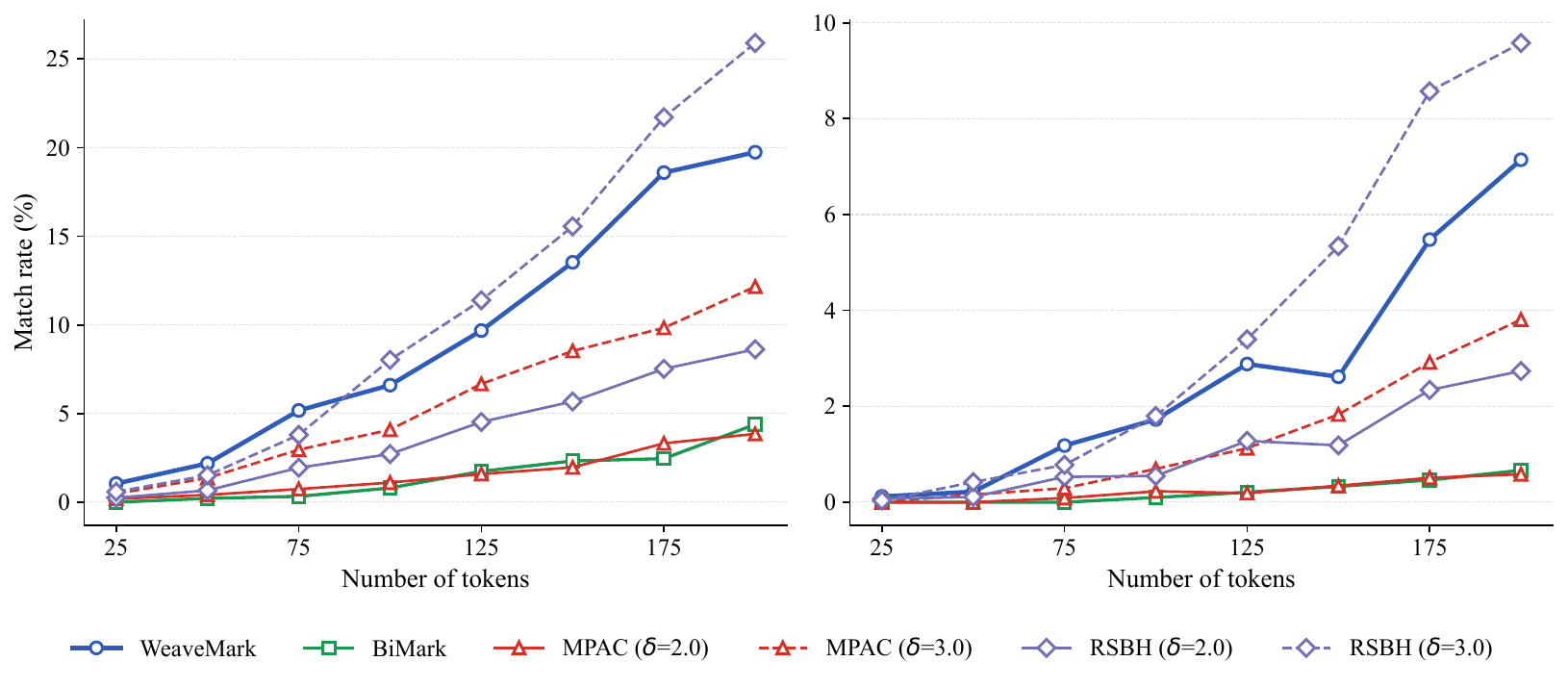}
    \caption{Robustness to stronger synonym substitution attacks with $\rho=0.3$.}
    \label{fig:robustness_extended}
\end{figure}

Figure~\ref{fig:robustness_extended} extends the robustness experiments in Section~\ref{sec:extraction} to a higher substitution ratio of $\rho=0.3$ for 12-bit and 16-bit messages. RSBH ($\delta=3.0$) outperforms \textsc{WeaveMark} at longer token lengths: at 200 tokens, it achieves 25.90\% versus 19.74\% for 12-bit messages and 9.58\% versus 7.14\% for 16-bit messages.

This contrasts with the results at $\rho=0.1$ and $\rho=0.2$, where \textsc{WeaveMark} consistently outperforms all baselines under the original method-specific settings. RSBH's advantage at $\rho=0.3$ comes at the cost of degraded text quality from stronger biased reweighting (Section~\ref{sec:quality}). Its smaller context window ($h=1$, compared with $h=2$ for \textsc{WeaveMark}) also limits the propagation of substitution errors. Appendix~\ref{app:window} discusses this effect and additional comparisons with a shared context window.

Table~\ref{tab:robustness_long} extends the evaluation to 24-bit and 32-bit messages with 200 generated tokens across substitution ratios $\rho=0.1, 0.2, 0.3$. At $\rho=0.1$ and $\rho=0.2$, \textsc{WeaveMark} consistently outperforms all baselines, including RSBH ($\delta=3.0$). At $\rho=0.3$, match rates of all methods drop to near zero, leaving little room for meaningful comparison.

\begin{table}[h]
\centering
\caption{Match rate (\%) at 200 tokens for 24-bit and 32-bit messages across substitution ratios.}
\label{tab:robustness_long}
\setlength{\tabcolsep}{6pt}
\renewcommand{\arraystretch}{1.2}
\small
\begin{tabular}{lcccccc}
\toprule
& \multicolumn{3}{c}{24-bit} & \multicolumn{3}{c}{32-bit} \\
\cmidrule(lr){2-4} \cmidrule(lr){5-7}
Method & $\rho{=}0.1$ & $\rho{=}0.2$ & $\rho{=}0.3$ & $\rho{=}0.1$ & $\rho{=}0.2$ & $\rho{=}0.3$ \\
\midrule
\textsc{WeaveMark}         & \textbf{60.13} & \textbf{14.41} & 0.72          & \textbf{32.10} & \textbf{2.72} & 0    \\
BiMark           & 4.95           & 0.24           & 0             & 0.68           & 0.06          & 0    \\
MPAC ($\delta{=}2.0$) & 0.59      & 0.11           & 0             & 0              & 0             & 0    \\
MPAC ($\delta{=}3.0$) & 4.56      & 0.80           & 0.15          & 0.49           & 0.05          & 0    \\
RSBH ($\delta{=}2.0$) & 11.18     & 1.68           & 0.26          & 2.38           & 0.30          & 0    \\
RSBH ($\delta{=}3.0$) & 45.34     & 10.13          & \textbf{1.04} & 22.03          & 1.79          & \textbf{0.04} \\
\bottomrule
\end{tabular}
\end{table}

\subsection{Insertion, deletion, and mixed edits}
\label{app:attacks}

Section~\ref{sec:extraction} evaluates synonym substitution, which preserves token positions. Table~\ref{tab:attacks_edit} reports match rates under insertion and deletion, which shift the positions of all subsequent tokens, and under mixed edits that apply insertion, deletion, and substitution at equal rates. \textsc{WeaveMark} retains 84.5\% at 20\% insertion and 73.0\% at 20\% deletion, whereas no baseline exceeds 26.2\% and 15.5\%, respectively.

\begin{table}[h]
\centering
\caption{Match rate (\%) under position-changing edits, for 16-bit messages with 200 generated tokens. Mixed edits apply insertion, deletion, and substitution at equal rates.}
\label{tab:attacks_edit}
\small
\setlength{\tabcolsep}{4pt}
\renewcommand{\arraystretch}{1.15}
\begin{tabular*}{\textwidth}{@{\extracolsep{\fill}}lcccccccccc@{}}
\toprule
& & \multicolumn{3}{c}{Insertion} & \multicolumn{3}{c}{Deletion} & \multicolumn{3}{c}{Mixed} \\
\cmidrule(lr){3-5}\cmidrule(lr){6-8}\cmidrule(lr){9-11}
Method & Clean & 10\% & 15\% & 20\% & 10\% & 15\% & 20\% & 10\% & 15\% & 20\% \\
\midrule
\textsc{WeaveMark} & \textbf{99.0} & \textbf{96.2} & \textbf{91.6} & \textbf{84.5} & \textbf{94.7} & \textbf{88.5} & \textbf{73.0} & \textbf{94.1} & \textbf{87.2} & \textbf{77.4} \\
BiMark     & 77.5 & 48.7 & 36.3 & 26.2 & 44.5 & 27.5 & 14.4 & 45.6 & 28.9 & 19.8 \\
RSBH       & 73.9 & 40.0 & 27.0 & 16.8 & 36.1 & 21.9 & 10.2 & 37.1 & 20.0 & 12.6 \\
MPAC       & 66.1 & 43.0 & 30.6 & 23.1 & 40.9 & 25.8 & 15.5 & 38.5 & 26.9 & 17.0 \\
XMark      & 54.1 & 25.5 & 16.1 & 9.9  & 23.9 & 11.3 & 6.6  & 23.1 & 12.5 & 7.7  \\
StealthInk & 31.0 & 9.0  & 5.6  & 2.9  & 9.0  & 4.4  & 2.2  & 7.5  & 3.4  & 2.4  \\
\bottomrule
\end{tabular*}
\end{table}

\subsection{Truncation and copy-paste}
\label{app:spans}

Table~\ref{tab:attacks_practical} reports truncation and copy-paste attacks. Truncation and copy-paste preserve the local contexts within retained
watermarked spans except near their boundaries, allowing
\textsc{WeaveMark} to maintain high recovery rates.

\begin{table}[h]
\centering
\caption{Match rate (\%) under attacks that preserve contiguous spans (truncation and copy-paste), for 16-bit messages with 200 generated tokens.}
\label{tab:attacks_practical}
\small
\setlength{\tabcolsep}{4pt}
\renewcommand{\arraystretch}{1.15}
\begin{tabular*}{\textwidth}{@{\extracolsep{\fill}}lccccc@{}}
\toprule
& \multicolumn{2}{c}{Truncation} & \multicolumn{3}{c}{Copy-paste} \\
\cmidrule(lr){2-3}\cmidrule(lr){4-6}
Method & 10\% & 20\% & 10\% & 15\% & 20\% \\
\midrule
\textsc{WeaveMark} & \textbf{98.7} & \textbf{98.4} & \textbf{98.1} & \textbf{97.2} & \textbf{95.7} \\
BiMark & 71.9 & 65.6 & 66.0 & 56.6 & 49.8 \\
RSBH & 68.0 & 60.1 & 60.4 & 53.7 & 42.8 \\
MPAC & 59.4 & 52.9 & 56.1 & 49.9 & 42.8 \\
XMark & 48.5 & 41.6 & 40.5 & 34.3 & 28.7 \\
StealthInk & 20.0 & 17.3 & 15.3 & 13.0 & 9.7 \\
\bottomrule
\end{tabular*}
\end{table}

\subsection{Text rewriting}
\label{app:rewriting}

Table~\ref{tab:attacks_rewriting} reports summarization and round-trip translation attacks. Unlike truncation and copy-paste, summarization and round-trip translation rewrite the text, and both are far more damaging. Round-trip translation is the hardest setting for every method: \textsc{WeaveMark} recovers 36.6\% where the strongest baseline reaches 4.0\%.

\begin{table}[h]
\centering
\caption{Match rate (\%) under summarization and round-trip translation, for 16-bit messages with 200 generated tokens. Summarization ratios r30, r50, and r70 denote the fraction of the original length retained.}
\label{tab:attacks_rewriting}
\small
\setlength{\tabcolsep}{4pt}
\renewcommand{\arraystretch}{1.15}
\begin{tabular*}{\textwidth}{@{\extracolsep{\fill}}lcccc@{}}
\toprule
& \multicolumn{3}{c}{Summarization} & Round-trip \\
\cmidrule(lr){2-4}\cmidrule(lr){5-5}
Method & r30 & r50 & r70 & (en,fr,en) \\
\midrule
\textsc{WeaveMark} & \textbf{61.5} & \textbf{80.3} & \textbf{84.9} & \textbf{36.6} \\
BiMark & 7.5 & 21.0 & 35.3 & 3.4 \\
RSBH & 6.7 & 16.8 & 28.5 & 2.0 \\
MPAC & 9.9 & 19.8 & 28.8 & 4.0 \\
XMark & 4.5 & 10.1 & 18.2 & 2.0 \\
StealthInk & 1.7 & 3.9 & 6.6 & 0.4 \\
\bottomrule
\end{tabular*}
\end{table}

We additionally evaluate paraphrasing attacks using DIPPER~\citep{krishna2023paraphrasing}, which rewrites watermarked text with controllable lexical and order diversity (L, O). Higher values indicate stronger paraphrasing.

Table~\ref{tab:dipper} shows bit accuracy and match rate for 16-bit messages with 200 generated tokens under three paraphrasing settings. \textsc{WeaveMark} achieves substantially higher match rates than all baselines across all three settings. Under (20, 0), which only changes lexical choices, \textsc{WeaveMark} achieves 59.10\% match rate compared with 13.49\% for BiMark. Under (0, 20), which only changes word order, \textsc{WeaveMark} achieves 86.94\% versus 42.17\%. Even under the most aggressive (20, 20) setting that simultaneously changes both lexical and order diversity, \textsc{WeaveMark} maintains 49.02\% match rate, while BiMark drops to 9.80\%. These results show that the redundancy from coded payload spreading and soft-decision ECC decoding makes \textsc{WeaveMark} robust under paraphrasing attacks.

Experiments covering stronger substitution and longer messages, insertion, deletion, mixed edits, truncation, copy-paste, summarization, round-trip translation, and paraphrasing demonstrate that \textsc{WeaveMark} can maintain reliable message recovery with substantial margins in settings where competing methods largely fail.

\begin{table}[t]
\centering
\caption{Robustness under DIPPER paraphrasing (16-bit, 200 tokens). (L, O) denotes lexical and order diversity. Bit/Match: bit accuracy and match rate (\%).}
\label{tab:dipper}
\small
\setlength{\tabcolsep}{8pt}
\renewcommand{\arraystretch}{1.25}
\begin{tabular*}{\textwidth}{l@{\extracolsep{\fill}}cccccc}
\toprule
& \multicolumn{2}{c}{(L=20, O=0)} & \multicolumn{2}{c}{(L=0, O=20)} & \multicolumn{2}{c}{(L=20, O=20)} \\
\cmidrule(lr){2-3} \cmidrule(lr){4-5} \cmidrule(lr){6-7}
Method & Bit & Match & Bit & Match & Bit & Match \\
\midrule
\textsc{WeaveMark}  & 80.78 & \textbf{59.10} & 89.43 & \textbf{86.94} & 78.18 & \textbf{49.02} \\
BiMark               & 84.33 & 13.49          & 92.29 & 42.17          & 81.80 & 9.80 \\
MPAC ($\delta{=}2.0$) & 75.25 & 3.56           & 81.41 & 9.93           & 73.70 & 2.44 \\
MPAC ($\delta{=}3.0$) & 82.89 & 12.98          & 89.33 & 31.98          & 80.84 & 9.67 \\
RSBH ($\delta{=}2.0$) & 71.43 & 12.61          & 80.63 & 36.47          & 69.60 & 9.96 \\
RSBH ($\delta{=}3.0$) & 77.87 & 28.43          & 88.10 & 59.35          & 76.15 & 22.88 \\
\bottomrule
\end{tabular*}
\end{table}

\section{Effect of the context window size}
\label{app:window}

The context window size $h$ determines how many preceding tokens seed the pseudorandom functions, and it affects extraction in two opposing ways. A smaller $h$ makes repeated contexts more likely; since unbiased reweighting requires that no context be reused for watermarking, more tokens are skipped and carry no watermark. A larger $h$ makes contexts more unique, but widens the span affected by an edit. For example, a substitution at position $t$ changes the observed token at that position and can alter the context-derived seeds at positions $t+1,\ldots,t+h$.

To quantify this trade-off, we vary $h \in \{2, 3, 4\}$ and measure the valid embedding ratio---the fraction of generated tokens that actually carry the watermark---together with match rates under the attacks of Appendix~\ref{app:robustness}. Table~\ref{tab:window_size} reports the results for 16-bit messages with 200 generated tokens.

\begin{table}[h]
\centering
\caption{Effect of the context window size $h$ on the valid embedding ratio and match rate (\%), for 16-bit messages with 200 generated tokens on Mistral-7B with OpenGen. Left: scattered edits, whose impact increases with $h$. Right: attacks that preserve contiguous spans, which are largely insensitive to $h$, together with summarization.}
\label{tab:window_size}
\small
\setlength{\tabcolsep}{5pt}
\renewcommand{\arraystretch}{1.15}
\begin{minipage}[t]{0.53\textwidth}
\centering
\begin{tabular}{lccc}
\toprule
& $h=2$ & $h=3$ & $h=4$ \\
\midrule
Valid embedding ratio & 86.6 & 94.0 & \textbf{97.2} \\
Clean text            & 99.0 & \textbf{99.2} & 99.1 \\
\midrule
Subst. (10\%)         & \textbf{81.8} & 74.6 & 59.3 \\
Subst. (20\%)         & \textbf{40.7} & 21.4 & 9.9  \\
Ins. (10\%)           & \textbf{96.2} & 94.6 & 89.7 \\
Ins. (20\%)           & \textbf{84.5} & 69.7 & 47.5 \\
Del. (10\%)           & \textbf{94.7} & 92.4 & 85.6 \\
Del. (20\%)           & \textbf{73.0} & 50.7 & 23.7 \\
Mix. (10\%)           & \textbf{94.1} & 92.8 & 87.4 \\
Mix. (20\%)           & \textbf{77.4} & 60.2 & 33.4 \\
\bottomrule
\end{tabular}
\end{minipage}\hfill
\begin{minipage}[t]{0.45\textwidth}
\centering
\begin{tabular}{lccc}
\toprule
& $h=2$ & $h=3$ & $h=4$ \\
\midrule
Trunc. (10\%)     & 98.7 & \textbf{99.1} & 98.9 \\
Trunc. (20\%)     & 98.4 & \textbf{98.9} & 98.8 \\
Copy-paste (10\%) & 98.1 & \textbf{98.7} & 98.6 \\
Copy-paste (20\%) & 95.7 & 96.9 & \textbf{97.4} \\
\midrule
Summ. (r50)       & 80.3 & \textbf{81.2} & 75.3 \\
Summ. (r70)       & \textbf{84.9} & 84.5 & 80.2 \\
\bottomrule
\end{tabular}
\end{minipage}
\end{table}

The valid embedding ratio rises from 86.6\% at $h=2$ to 97.2\% at $h=4$, yet the clean match rate is essentially unchanged (99.0\% versus 99.1\%). At 200 tokens, payload spreading already supplies enough observations per bit that the skipped tokens do not limit recovery, so the ratio matters only when tokens are scarce relative to the message length.

Under attack, the two groups of edits behave differently. Attacks that scatter independent edits across the text---substitution, insertion, deletion, and their mixture---degrade sharply as $h$ grows, because a larger context window allows each edit to disrupt extraction over a wider span. At 20\% substitution, the match rate falls from 40.7\% at $h=2$ to 9.9\% at $h=4$. Attacks that preserve contiguous spans---truncation and copy-paste---are largely insensitive to $h$ in these experiments, since their effects on context reconstruction remain localized near span boundaries. Summarization changes little between $h=2$ and $h=3$ and declines at $h=4$, a far smaller effect than for scattered edits.

We therefore use $h=2$ throughout: it is the strongest setting under every scattered-edit attack while losing nothing on clean text.

\textbf{Context window size and substitution robustness.}
The comparison at $\rho=0.3$ in Appendix~\ref{app:substitution} also reflects differences in context window size. \textsc{WeaveMark} and BiMark use $h=2$, whereas MPAC and RSBH use $h=1$. A substitution can therefore disrupt extraction at three positions for \textsc{WeaveMark}, compared with two for RSBH, giving RSBH an advantage under substitution attacks.

For \textsc{WeaveMark} and BiMark, a larger context window reduces repeated contexts, which must be skipped to preserve unbiasedness. This retains more tokens for watermark embedding but makes extraction more sensitive to edits. MPAC and RSBH do not require this restriction and use their recommended window size of $h=1$. 
Under the original method-specific settings, \textsc{WeaveMark} already achieves superior message recovery across a broad range of conditions, with exceptions under severe substitution ($\rho=0.3$). The additional comparisons with a shared window of $h=2$ further demonstrate its robustness across diverse attacks (Tables~\ref{tab:attacks_edit},~\ref{tab:attacks_practical}, and~\ref{tab:attacks_rewriting}), where it again outperforms all evaluated baselines. Together, these results demonstrate strong robustness under both method-specific and shared context-window settings.

\section{Voting method comparison}
\label{app:voting}

When extracting an embedded bit, the per-token votes can be aggregated in three ways. 
\emph{Hard voting} compares $v_0[p]$ and $v_1[p]$ and outputs the majority bit; this discards confidence information, so a narrow majority such as $(100,99)$ is treated the same as a decisive majority such as $(100,0)$. 
\emph{Soft voting} uses the signed vote difference $v_1 - v_0$ as the per-bit score and decodes by maximum-likelihood correlation with the codebook. This preserves vote count but treats the magnitude of the difference as a linear measure of confidence---a difference of $+4$ from 4 unanimous votes $(4,0)$ is treated identically to $+4$ from 100 mostly-balanced votes $(52,48)$. \emph{LLR voting} uses the log-likelihood ratio $\log\bigl((v_1+0.5)/(v_0+0.5)\bigr)$ as the score, which primarily reflects the relative balance between the two vote counts.

\begin{figure}[H]
    \centering
    \includegraphics[width=\textwidth]{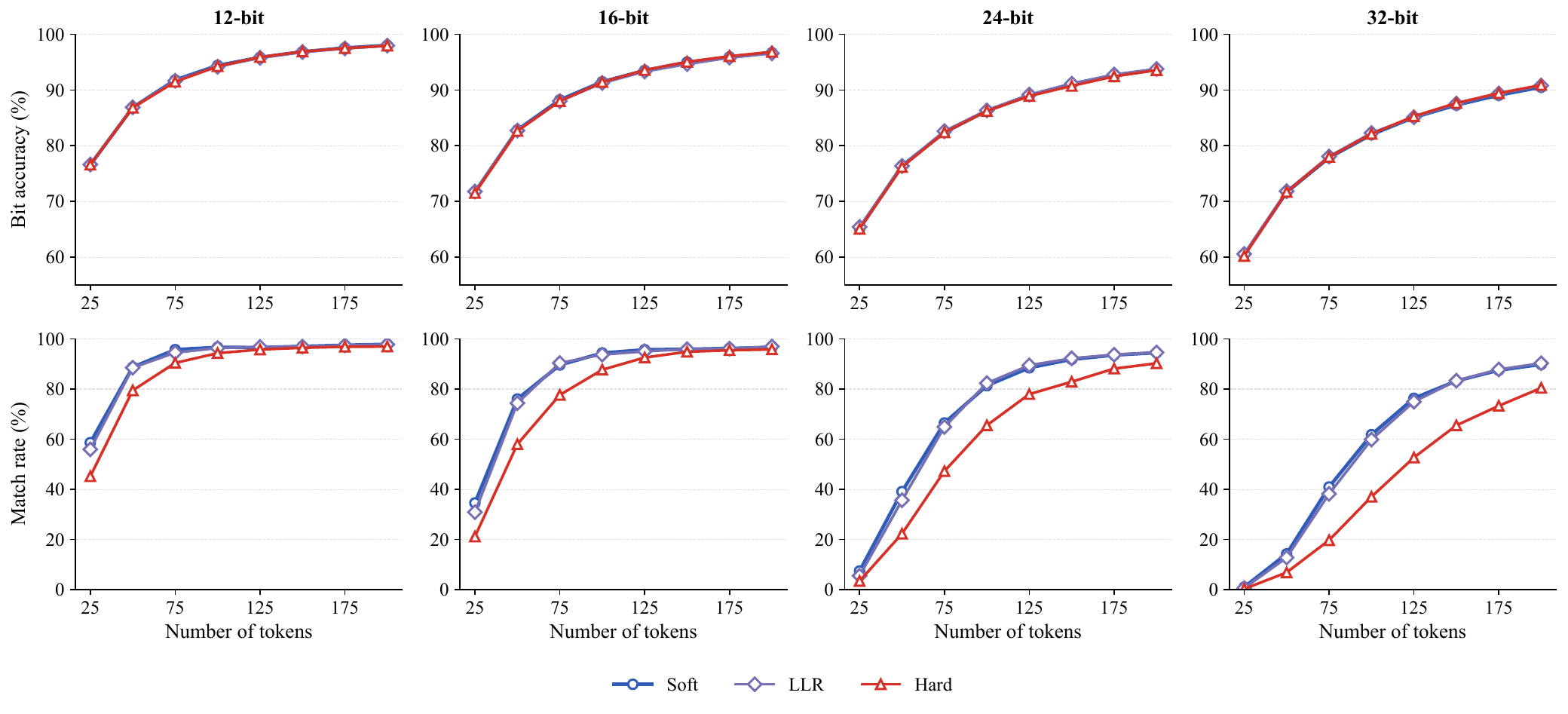}
    \caption{Comparison of hard, soft, and LLR voting across message lengths.}
    \label{fig:voting}
\end{figure}

Figure~\ref{fig:voting} reports bit accuracy and match rate for all three methods across message lengths and token counts. Hard voting consistently underperforms the other two methods on match rate at all lengths, with the gap widening as message length increases---at 32-bit, 100 tokens, hard voting achieves 37.06\% match rate while soft voting achieves 61.80\%. Soft voting and LLR voting are nearly indistinguishable at moderate to long lengths (125+ tokens), where they alternate within 1 percentage point. At shorter lengths (25--100 tokens), soft voting consistently outperforms LLR by 1--3 percentage points. 

With short texts, vote counts vary across bits. Soft voting retains the strength of the accumulated evidence, whereas the LLR score mainly reflects vote proportions and can overemphasize bits with few votes. This difference diminishes as longer texts provide more votes per bit. We therefore use soft voting in all main experiments.

Extracting richer side information from the watermarking process to improve soft inputs for ECC decoding is a particularly promising direction for future work.

\section{Perplexity analysis}
\label{app:perplexity}

\begin{figure}[H]
\centering
\begin{minipage}[t]{0.48\textwidth}
    \centering
    \includegraphics[width=\textwidth]{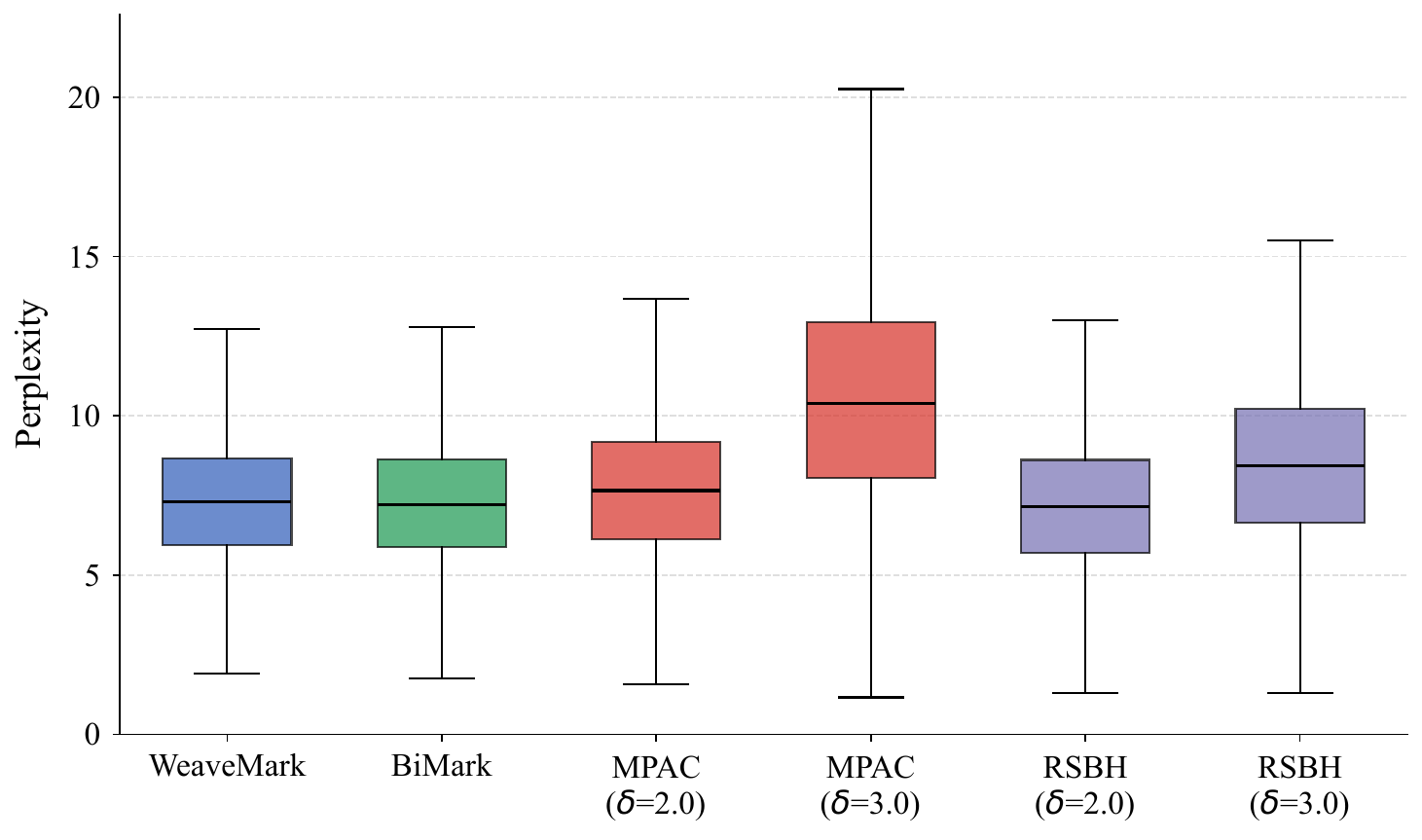}
    \caption{PPL distribution at 200 tokens (16-bit).}
    \label{fig:ppl_boxplot}
\end{minipage}
\hfill
\begin{minipage}[t]{0.48\textwidth}
    \centering
    \includegraphics[width=\textwidth]{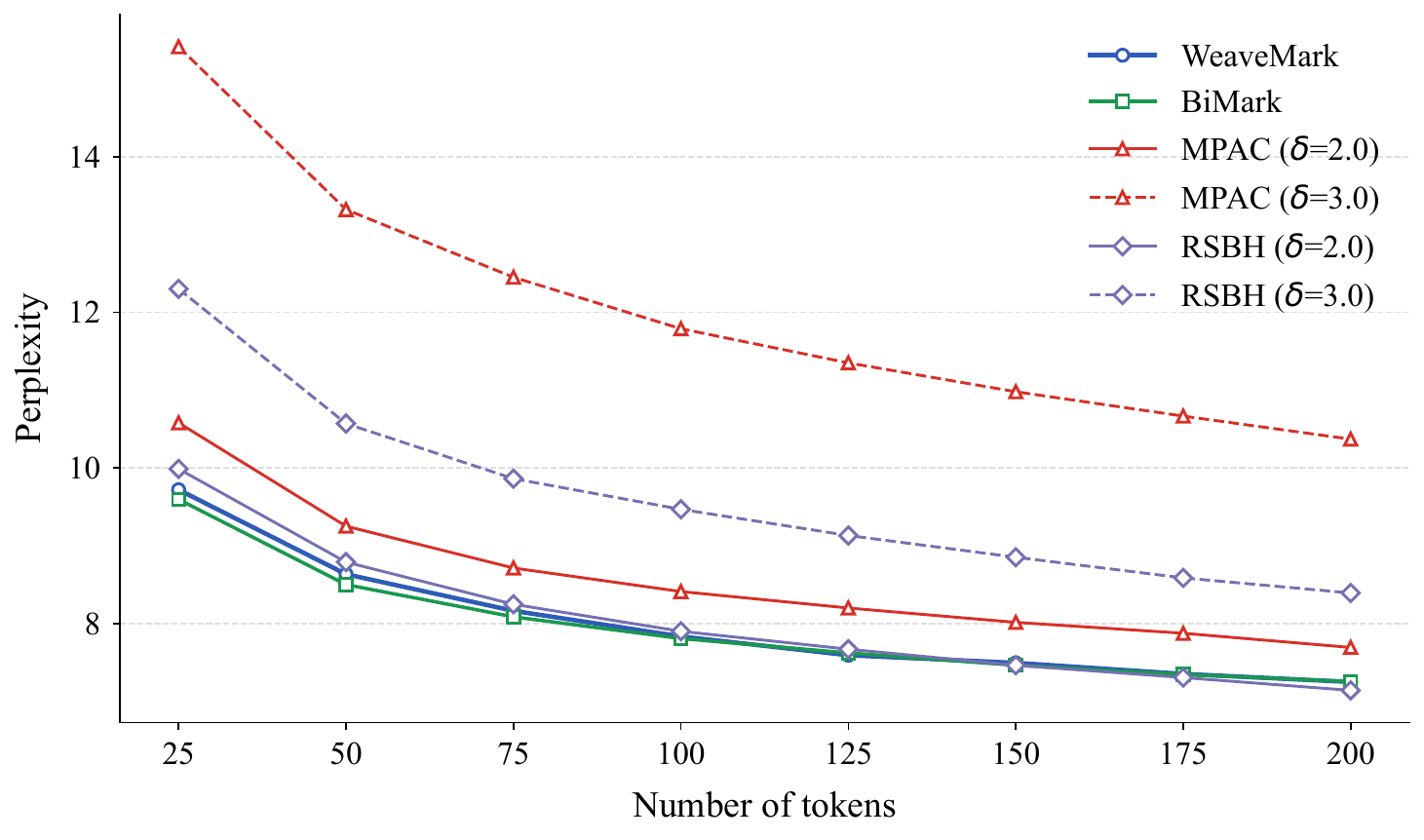}
    \caption{Mean PPL results (16-bit).}
    \label{fig:ppl_curve}
\end{minipage}
\end{figure}

We measure perplexity (PPL) with Gemma-2-9B~\citep{team2024gemma2} on watermarked text generated with LLaMA-3-8B on the C4 RealNewsLike dataset, using 16-bit messages. Figure~\ref{fig:ppl_boxplot} shows the PPL distribution at 200 tokens, and Figure~\ref{fig:ppl_curve} shows mean PPL across token lengths.

Across both figures, our method yields PPL comparable to BiMark, while MPAC and RSBH at $\delta=3.0$ show noticeably higher PPL---consistent with the downstream task results in Section~\ref{sec:quality}.

\textbf{Note on PPL measurement.}
With our generation setup---an 8B-parameter model under 4-bit quantization---the model occasionally produces repetitive patterns where it copies its own context. Such repetitions yield artificially low PPL since repeated tokens are highly predictable. This affects all methods, including non-watermarked text generation, so relative comparisons remain meaningful, but absolute PPL values should be interpreted with caution.

\section{Statistical reliability}
\label{app:stats}

This appendix addresses the statistical reliability of the reported results. Appendix~\ref{app:protocol} describes prompt construction, sample counts, and sampling uncertainty. Appendix~\ref{app:replication} examines how the match rate varies with the prompts, the messages, and the key. Appendix~\ref{app:threshold} covers the calibration of the FPR=1\% threshold used for zero-bit detection.

\subsection{Evaluation protocol}
\label{app:protocol}

A prompt is the first 100 words of a source document, cut back to the last sentence boundary it contains; documents with none are discarded. For each configuration we draw prompts until at least $3{,}000$ watermarked texts reach the target length, except in the replication study of Appendix~\ref{app:replication}. Reported results are computed over all retained texts, and all methods use the same prompt pool.

Let $N$ denote the actual number of evaluated texts. Under binomial sampling, a reported proportion with true rate $p$ has standard error $\sqrt{p(1-p)/N}$, bounded by $\sqrt{0.25/N}$. For $N\ge 3{,}000$, using $N=3{,}000$ gives conservative normal-approximation $95\%$ confidence half-widths: at most $1.79$ percentage points (pp), $1.07$~pp at a match rate of $90\%$, and $0.36$~pp at a TPR of $99\%$. For TPR, these calculations condition on a fixed threshold; threshold-calibration uncertainty is addressed in Appendix~\ref{app:threshold}.

\subsection{Sensitivity to prompts, messages, and keys}
\label{app:replication}

\textsc{WeaveMark} derives its watermark seeds from the preceding context and the secret key, so prompts are not the only source of randomness. Table~\ref{tab:replication} reports ten replications of the extraction experiment, each with a fresh prompt subset from the 7{,}711 OpenGen prompts, fresh messages, and a fresh key, at $1{,}500$ samples per replication. The observed standard deviations are no larger than the corresponding binomial sampling standard errors of $0.26$~pp and $0.82$~pp, indicating stable performance across the evaluated replications.

\begin{table}[h]
\centering
\caption{Match rate (\%) across ten replications with independently resampled prompts, messages, and secret keys, in the setting of Appendix~\ref{app:extended}. Confidence intervals are $t$-based over the ten replications.}
\label{tab:replication}
\small
\setlength{\tabcolsep}{6pt}
\renewcommand{\arraystretch}{1.15}
\begin{tabular*}{\textwidth}{@{\extracolsep{\fill}}lccc@{}}
\toprule
Setting & Mean & Std.\ dev. & 95\% CI \\
\midrule
16-bit, 200 tokens & 98.98 & 0.200 & [98.83, 99.12] \\
32-bit, 200 tokens & 88.49 & 0.660 & [88.02, 88.96] \\
\bottomrule
\end{tabular*}
\end{table}

\subsection{Setting the zero-bit detection threshold}
\label{app:threshold}

Zero-bit detection compares a text's score with a threshold. A lower threshold can increase the TPR on watermarked text, but also the FPR on non-watermarked text. For fair comparison, we calibrate each method's threshold under a common FPR limit of $1\%$.

\paragraph{Estimating the null distribution.}
The null distribution describes detection scores for non-watermarked text. We estimate it from a separate set of about $1{,}950$ texts per method, using prompts constructed as in Appendix~\ref{app:protocol}. One score per text leaves about twenty observations in the upper $1\%$ tail. We therefore re-score each text under $K=20$ independently sampled keys ($K=10$ for EXP-edit), yielding nearly $40{,}000$ scores per length, or half for EXP-edit. Texts are generated independently of these keys, providing valid null observations, although scores sharing a text may be correlated.

\paragraph{Selecting the threshold.}
Scores at or above the threshold indicate a watermark. The $99$th-percentile rule yields empirical FPRs of $1.01\%$--$2.19\%$ because scores tied at the threshold are admitted together. For each (method, length) pair, we select the lowest observed null score whose empirical calibration FPR does not exceed $1\%$. We assess TPR uncertainty using $1{,}000$ prompt-level cluster-bootstrap resamples, recalibrating on each resampled null to include threshold uncertainty.

\paragraph{Validating on held-out texts.}
We check whether calibrated thresholds generalize to held-out texts. Over $200$ random splits, we calibrate on half the texts and evaluate FPR on the rest, keeping each text's scores together. Mean held-out FPRs are at most $1.003\%$, supporting calibration in these settings (Table~\ref{tab:fpr_heldout}). Averages are $0.980\%$--$0.995\%$ for \textsc{WeaveMark}, EXP-edit, and SynthID, versus $0.806\%$ for KGW and $0.842\%$ for DiPmark.

\begin{table}[h]
\centering
\caption{Held-out FPR (\%), averaged over 200 random splits in which the threshold is computed on one half of the non-watermarked texts and evaluated on the other half. Columns give the number of generated tokens.}
\label{tab:fpr_heldout}
\tablefootnotesize
\setlength{\tabcolsep}{6pt}
\renewcommand{\arraystretch}{1.2}
\begin{tabular*}{\textwidth}{@{\extracolsep{\fill}}lcccccccc@{}}
\toprule
Method & 50 & 75 & 100 & 125 & 150 & 175 & 200 & Avg. \\
\midrule
\textsc{WeaveMark}      & 0.989 & 0.981 & 0.989 & 1.003 & 0.981 & 0.992 & 0.976 & 0.987 \\
EXP-edit                & 0.961 & 0.987 & 1.003 & 0.977 & 0.969 & 0.982 & 0.983 & 0.980 \\
SynthID                 & 0.986 & 0.997 & 1.002 & 0.992 & 0.998 & 1.000 & 0.987 & 0.995 \\
KGW ($\delta=2.0$)      & 0.764 & 0.883 & 0.792 & 0.655 & 0.876 & 0.775 & 0.895 & 0.806 \\
DiPmark ($\alpha=0.45$) & 0.789 & 0.903 & 0.720 & 0.854 & 0.911 & 0.827 & 0.888 & 0.842 \\
\bottomrule
\end{tabular*}
\end{table}

\paragraph{Accounting for discrete scores.}
KGW and DiPmark leave some FPR budget unused, potentially reducing TPR. Their statistics depend on integer counts and take only a few dozen distinct null values, versus thousands for the other methods. Lowering the threshold by one step increases empirical FPR by up to $1.15$~pp and $0.56$~pp, respectively, versus at most $0.13$~pp elsewhere, making it difficult to approach $1\%$ without exceeding it.

To compare methods using the full budget, we interpolate between the selected and next lower thresholds, denoted by operating points $(\mathrm{fpr}_c,\mathrm{tpr}_c)$ and $(\mathrm{fpr}_r,\mathrm{tpr}_r)$, with $\mathrm{fpr}_c\le 0.01<\mathrm{fpr}_r$. Choosing the lower threshold with probability $w$ and the selected threshold otherwise gives
\[
w=\frac{0.01-\mathrm{fpr}_c}{\mathrm{fpr}_r-\mathrm{fpr}_c},
\qquad
\mathrm{tpr}_{\mathrm{interp}}=\mathrm{tpr}_c+w(\mathrm{tpr}_r-\mathrm{tpr}_c).
\]
This mixture has an expected empirical FPR of exactly $1\%$ on the calibration set. Its TPR is the weighted average of the two TPRs.

Interpolation raises average TPR by $0.05$~pp for KGW and $2.11$~pp for DiPmark, and changes the other methods' averages by at most $0.01$~pp (Table~\ref{tab:interp_tpr}). The ranking in Table~\ref{tab:zerobit_compare} remains unchanged, with DiPmark still $17.6$~pp below \textsc{WeaveMark} on average.

\begin{table}[h]
\centering
\caption{Clean zero-bit TPR (\%) under the selected threshold and under interpolation to an expected empirical FPR of 1\% on the calibration set. The selected rows are the values reported in Table~\ref{tab:zerobit_compare}. Columns give the number of generated tokens.}
\label{tab:interp_tpr}
\tablefootnotesize
\setlength{\tabcolsep}{5pt}
\renewcommand{\arraystretch}{1.2}
\begin{tabular*}{\textwidth}{@{\extracolsep{\fill}}llcccccccc@{}}
\toprule
Method & Threshold & 50 & 75 & 100 & 125 & 150 & 175 & 200 & Avg. \\
\midrule
\multirow{2}{*}{\textsc{WeaveMark}} & Selected     & 97.94 & 98.53 & 98.67 & 98.45 & 98.81 & 98.99 & 98.95 & 98.62 \\
                                    & Interpolated & 97.95 & 98.53 & 98.67 & 98.45 & 98.81 & 98.99 & 98.95 & 98.62 \\
\midrule
\multirow{2}{*}{EXP-edit}           & Selected     & 99.40 & 99.60 & 99.63 & 99.53 & 99.77 & 99.80 & 99.93 & 99.67 \\
                                    & Interpolated & 99.40 & 99.61 & 99.63 & 99.53 & 99.77 & 99.83 & 99.93 & 99.67 \\
\midrule
\multirow{2}{*}{SynthID}            & Selected     & 98.09 & 98.41 & 98.61 & 98.68 & 98.92 & 99.08 & 99.26 & 98.72 \\
                                    & Interpolated & 98.09 & 98.41 & 98.61 & 98.68 & 98.92 & 99.08 & 99.26 & 98.72 \\
\midrule
\multirow{2}{*}{KGW ($\delta=2.0$)} & Selected     & 92.98 & 96.74 & 97.49 & 97.67 & 98.10 & 98.61 & 99.13 & 97.25 \\
                                    & Interpolated & 93.01 & 96.78 & 97.63 & 97.78 & 98.12 & 98.64 & 99.13 & 97.30 \\
\midrule
\multirow{2}{*}{DiPmark ($\alpha=0.45$)} & Selected     & 46.43 & 72.21 & 78.29 & 85.61 & 88.46 & 89.70 & 91.60 & 78.90 \\
                                    & Interpolated & 54.33 & 73.61 & 81.18 & 86.27 & 88.46 & 90.78 & 92.42 & 81.01 \\
\bottomrule
\end{tabular*}
\end{table}
\end{document}